\documentclass[fleqn,12pt]{article}
\usepackage{natbib}
\usepackage{amsthm,amsfonts,amsopn,amsmath,amssymb,vmargin,verbatim,setspace}
\usepackage{pgf,pgfplots,tikz}
\usepackage{booktabs}

\RequirePackage[pagebackref=true]{hyperref}
\RequirePackage{hypernat}
\usepackage{float}

\usetikzlibrary{arrows,calc,fit,matrix,positioning,shapes.multipart,shapes.symbols}
\hypersetup{colorlinks=true,linkcolor=cyan,urlcolor=cyan,citecolor=darkgray,
    pdftitle=The Likelihood of Mixed Hitting Times,
    pdfauthor=Jaap Abbring and Tim Salimans,
    pdfsubject=JEL Codes C14 C41,
    pdfkeywords=duration analysis first passage time identification Laplace transform Levy process maximum likelihood Mellin inverse formula mixture optimal stopping strike duration,
    pdfdisplaydoctitle=true}

\setpapersize{A4} \setmargrb{25mm}{18mm}{25mm}{23mm}
\doublespace

\renewenvironment{thebibliography}[1]{\begin{oldthebibliography}{#1}\setlength{\itemsep}{4.2pt}}{\end{oldthebibliography}}

\newcommand{\E}{\mathbb{E}}

\newcommand{\R}{\mathbb{R}}
\newcommand{\Rp}{[0,\infty)}
\newcommand{\Rpp}{(0,\infty)}
\newcommand{\Rpext}{[0,\infty]}

\newcommand{\C}{\mathbb{C}}

\renewcommand\L{{\cal G}}
\newcommand\LT{{\cal F}}

\newcommand\realint{q}
\newcommand\pathconst{c}
\newcommand{\epsin}{\delta}
\newcommand{\epsout}{\epsilon}

\theoremstyle{plain}
\newtheorem{theorem}{Theorem}
\newtheorem{lemma}{Lemma}
\newtheorem{corollary}{Corollary}

\theoremstyle{definition}
\newtheorem{assumption}{Assumption}
\newtheorem{definition}{Definition}

\newcommand{\cites}[1]{\citeauthor{#1}'s (\citeyear{#1})}

\begin{document}

\title{The Likelihood of Mixed Hitting Times\thanks{Forthcoming in the {\em Journal of Econometrics}: \href{https://doi.org/10.1016/j.jeconom.2019.08.017}{doi.org/10.1016/j.jeconom.2019.08.017}.}}

\author{Jaap H. Abbring\thanks{%
Department of Econometrics \& OR, Tilburg University, P.O. Box 90153, 5000 LE Tilburg, The Netherlands; and CEPR. E-mail:
\href{mailto:jaap@abbring.org}{jaap@abbring.org}. Web: \href{http://jaap.abbring.org}{jaap.abbring.org}.}\and Tim Salimans\thanks{Brain Team, Google Research, Amsterdam, The Netherlands. E-mail:
\href{mailto:salimans@google.com}{salimans@google.com}. Web: \href{http://github.com/TimSalimans}{github.com/TimSalimans}.
\newline\newline 
\copyright~2021. This manuscript  is made available under a \href{http://creativecommons.org/licenses/by-nc-nd/4.0/}{CC BY-NC-ND 4.0 International license}. 
}}

\date{April 2021}

\maketitle

\begin{abstract}
We present a method for computing the likelihood of a mixed hitting-time model that specifies durations as the first time a latent L\'{e}vy process crosses a heterogeneous threshold. This likelihood is not generally known in closed form, but its Laplace transform is. Our approach to its computation relies on numerical methods for inverting Laplace transforms that exploit special properties of the first passage times of L\'{e}vy processes. We use our method to implement a maximum likelihood estimator of the mixed hitting-time model in MATLAB. We illustrate the application of this estimator with an analysis of \cites{jem85:kennan} strike data. 

\end{abstract}

\thispagestyle{empty}


\newpage

\section{Introduction}

Mixed hitting-time (MHT) models are mixture duration models that specify durations as the first time a latent stochastic process crosses a heterogeneous threshold. They are of substantial interest because they can be applied to the analysis of optimal stopping decisions by heterogeneous agents \citep{ecta12:abbring,are10:abbring}. In particular, they can be applied to problems that do not lead to the mixed proportional hazards (MPH) model, \cites{ecta79:lancaster} and \cites{demo79:vaupeletal} popular extension of the \cite{jrssb72:cox} proportional hazards model. Examples include models of job durations, marriage durations, and the entry and exit of firms that are driven by Brownian motions and more general persistent processes. Hitting-time duration models are also popular in statistics for their structural and descriptive appeal \citep{ss06:leewhitmore}.

This paper considers likelihood-based empirical methods for an MHT model in which the latent process is a spectrally-negative L\'{e}vy process, a continuous-time process with stationary and independent increments and no positive jumps, and the threshold is proportional in the effects of observed regressors and unobserved heterogeneity. Spectrally-negative L\'{e}vy processes include Brownian motions with linear drifts and Poisson processes compounded with negative shocks as well-known special cases. Following empirical practice with mixture duration models such as the mixed proportional hazards model, we focus on parametric MHT models, and propose flexible parameterizations that can approximate arbitrary functional forms by increasing the number of parameters. The main obstacle in applying standard parametric likelihood methods is that, in general, we have no explicit expression for the MHT model's likelihood. However, an explicit expression for its Laplace transform is always available. Our approach to likelihood computation exploits this.

We focus on the case in which the latent L\'{e}vy process has a nontrivial Gaussian component. We first show that this ensures that the model implies a duration distribution with nonzero Lebesgue density at all positive durations and that it is nonparametrically identified up to innocuous scale normalizations. We then adapt numerical methods for the inversion of the Laplace transforms of the hitting times of L\'{e}vy processes with nontrivial Gaussian components to compute the conditional density and survival function implied by the MHT model. In turn, these are used to construct a likelihood for independently censored duration data. If the latent process is a Brownian motion, the likelihood can be explicitly expressed in terms of mixed inverse Gaussian densities and survival functions. Therefore, we can use this special case as a benchmark for evaluating the quality of our procedure for computing the likelihood. We show that the numerical inversion that is required in the general case is sufficiently fast and precise to make maximum likelihood estimation feasible even if no explicit expression of the likelihood is available.

We implement a maximum likelihood estimator that uses this computational strategy in MATLAB, and illustrate its application with a  reconsideration of \cites{jem85:kennan} empirical analysis of US contract strike durations.\footnote{We provide MATLAB code that implements the methods in this paper in a  public repository at \href{https://github.com/jabbring/mht-likelihood}{github.com/jabbring/mht-likelihood}. The results in this paper can be replicated by running {\tt make} in version v1.1.1 of this code, which we have deposited as \cite{zenodo21:abbringsalimans}.} Our strategy for computing the MHT model's likelihood can also be used to implement other likelihood-based empirical methods. For example, it can be combined with data augmentation and Markov chain Monte Carlo techniques to implement Bayesian estimators of the MHT model.  

\cite{ecta12:abbring} presented the MHT model studied in this paper, analyzed its empirical content, and highlighted its close relation to optimal stopping problems in economics. This paper shows that the restriction to an MHT model with a nontrivial Gaussian component suffices for its identification. It operationalizes this model by providing and analyzing feasible methods for computing its likelihood and its maximum likelihood estimator. 

\cite{singleton2001estimation} developed similar methods for a different class of models, discretely sampled affine diffusions. He noted that the density of an observation of such a diffusion conditional on the previous observation is not known explicitly, but that its characteristic function is. He proposed a maximum likelihood estimator based on the Fourier inverse of this characteristic function. This paper's methods for the MHT model instead rely on the inversion of Laplace transforms  and exploit specific results for the first passage times of L\'{e}vy processes.

Alternatively, we could avoid computation of the likelihood altogether by constructing an estimator directly from the equality of the Laplace transform of the duration data implied by the true model and its empirical analog. \citet[][Section 5.3]{ecta12:abbring} sketched such a generalized method of moments (GMM) estimator for the MHT model. A disadvantage of this alternative approach is that, unlike this paper's likelihood-based approach, it cannot straightforwardly handle censored duration data because we only have an expression of the Laplace transform of the complete (uncensored) duration distribution.\footnote{\citet{singleton2001estimation} developed a similar GMM estimator for discretely sampled diffusions, based on their characteristic function. In that context, censoring is not important and such a GMM estimator is a natural alternative to maximum likelihood.} Moreover, a practical implementation of such a GMM estimator is generally less efficient than maximum likelihood. Therefore, this paper focuses on likelihood-based methods. 

The remainder of this paper is organized as follows. Section \ref{s:model} reviews the MHT model and the corresponding characterization of the data presented in \cite{ecta12:abbring}. It also introduces the assumption that the latent process has a nontrivial Gaussian component and explores its implications, including novel nonparametric and parametric identification results. Section \ref{s:ML} presents a method for the computation of the model's log likelihood and its derivatives and discusses maximum likelihood estimation. Section \ref{s:experiments} assesses the numerical accuracy of our method and Section \ref{s:strike} applies it to strike data. Section \ref{s:concl} briefly discusses extensions to Bayesian and sieve estimators and reviews possible applications.


\section{Mixed Hitting-Time Model}
\label{s:model}

\subsection{Specification}

Following \citet[][Section 2]{ecta12:abbring}, we model the distribution of a random duration $T$ conditional on
observed covariates $X$ by specifying $T$ as the first time a
real-valued L\'{e}vy process $\{Y\}\equiv\{Y(t); t\geq 0\}$ crosses
a threshold that depends on $X$ and some unobservables $V$; assuming that $\{Y\}$, $X$, 
and $V$ are mutually independent; and specifying a marginal distribution of $V$.

A L\'{e}vy process is the continuous-time equivalent of a random
walk: It has stationary and independent increments. \citet{cup96:bertoin} provides a comprehensive analysis of
L\'{e}vy processes. Formally, we have
\begin{definition}
\label{def:levy} A {\em L\'{e}vy process} is a stochastic process
$\{Y\}$ such that the increment $Y(t+\Delta)-Y(t)$ is independent of
$\{Y(\tau); 0\leq\tau\leq t\}$ and has the same distribution as
$Y(\Delta)$, for every $t,\Delta\geq 0$.
\end{definition}

\noindent We take $\{Y\}$ to have right-continuous sample paths with
left limits. Note that Definition \ref{def:levy} implies that
$Y(0)=0$ almost surely.

An important example of a L\'{e}vy process is the scalar Brownian
motion with drift, in which case $Y(\Delta)$ is normally distributed
with mean $\mu\Delta$ and variance $\sigma^2\Delta$, for some scalar
parameters $\mu\in\R$ and $\sigma\in\Rp$. The Brownian motion is the
single L\'{e}vy process with continuous sample paths. In general,
L\'{e}vy processes may have jumps. Examples are compound Poisson
processes, which have independently and identically distributed
jumps at Poisson times. More generally, the jump process $\{\Delta
Y\}$ of a L\'{e}vy process $\{Y\}$ is a Poisson point process with
characteristic measure $\Upsilon$ such that
$\int\min\{1,y^2\}\Upsilon(dy)<\infty$, and any L\'{e}vy process
$\{Y\}$ can be written as the sum of a Brownian motion with drift
and an independent pure-jump process with jumps governed by such a
point process \citep[][Chapter I, Theorem 1]{cup96:bertoin}. The
characteristic measure of $\{Y\}$'s jump process is called its {\em
L\'{e}vy measure} and, together with the drift and dispersion
parameters of its Brownian motion component, fully characterizes
$\{Y\}$'s distributional properties.

Throughout the paper, we will focus on spectrally-negative L\'{e}vy
processes, which are L\'{e}vy processes of which the characteristic
measure $\Upsilon$ has negative support, i.e. L\'{e}vy
processes without positive jumps.  This greatly facilitates the
analysis of their hitting times, because it excludes that they jump across the threshold.
Let $\{Y\}$ be a spectrally-negative L\'{e}vy process and  $T(y)\equiv\inf\{t\geq 0:Y(t)>y\}$
the first time it hits a threshold $y\in\Rp$. Here, we use the convention that $\inf\emptyset\equiv\infty$;
that is, we set $T(y)=\infty$ if $\{Y\}$ never crosses $y$, which happens with positive probability for some specifications of $\{Y\}$. We exclude the trivial case that $\{Y\}$ is weakly decreasing and $T(y)=\infty$ almost surely.\footnote{This is implied by Assumption \ref{ass:sigma}, which we will introduce only later because it is easier to formulate after developing the model's characterization (which requires the weaker assumption made here).} 

Denote the support of the observed covariates $X$ with ${\cal X}\subseteq\R^K$, let $V$ have distribution $G$ on $\Rpp$, and recall that $\{Y\}$, $X$, and $V$ are mutually independent. The (proportional) mixed hitting-time (MHT) model specifies the cumulative distribution $F(\cdot|x,v)$ of $T$ conditional on $(X,V)=(x,v)\in{\cal X}\times\Rpp$ as $F(t|x,v)=\Pr\left[T(\phi(x)v)\leq t\right]$, for some measurable function $\phi:{\cal X}\rightarrow\Rpp$.\footnote{For expositional convenience, we have restricted the supports of $\phi(X)$ and $V$, and therefore of the threshold $\phi(X)V$, to $\Rpp$. It is straightforward to extend the analysis to $\Rpext$-valued thresholds, as in \citet[][Section 2.2 and Appendix A]{ecta12:abbring}. This would allow for a probability mass at zero duration (as $T(0)=0$ almost surely) and, with $T(\infty)\equiv\infty$, a mass of ``stayers.''}  Integrating out $v$ with respect to the distribution $G$ of $V$ gives the distribution  $F(t |x)=\int F(t|x,v) dG(v)=\int \Pr\left[T(\phi(x)v)\leq t\right]dG(v)$ of $T|X=x$. We note the corresponding ``survival function'' with $\overline F(t|x)\equiv 1-F(t|x)$.

\subsection{Characterization}
\label{ss:char}

The distribution $F(\cdot|x,v)$ is fully determined by its Laplace transform,
$\LT(s|x,v)\equiv\int_{\Rp}\exp\left(-s t\right)dF(t|x,v)$, $s\in\Rp$. Note that 
$\LT(0|x,v)=\lim_{t\rightarrow\infty}F(t|x,v)$ may be smaller than 1 if $\{Y\}$ is such that, with positive probability, it never hits $\phi(x)v$. 

\citet[][Section 4.1]{ecta12:abbring} showed that the Laplace transform $\LT(\cdot|x,v)$, unlike $F(\cdot|x,v)$ itself, can be explicitly given for any specification of the latent process $\{Y\}$. This first requires a common probabilistic characterization of $\{Y\}$, in terms of its characteristic function. \citet[][Section VII.1]{cup96:bertoin} shows that
$\E\left[\exp\left(s Y(t)\right)\right]=\exp\left[\psi(s)t\right]$,
for all $s\in\C$ with real part $\Re\,s\geq 0$, with the {\em Laplace exponent} $\psi$ given
by the L\'{e}vy-Khintchine formula,
\begin{equation}
\label{eq:levykitch}
\psi(s)=\tilde\mu s+\frac{\sigma^2}{2}
s^2+\int_{(-\infty,0)}\left\{\mathrm{e}^{s y}-1-s y
I(y>-1)\right\}\Upsilon(d y).
\end{equation}

\noindent Here, $I(\cdot)\equiv 1$ if $\cdot$ is true and $0$ otherwise, $\tilde\mu\in\mathbb{R}$ absorbs any linear drift of $\{Y\}$, $\sigma\geq 0$ is the dispersion parameter of its Brownian motion component; and
$\Upsilon$  is the L\'{e}vy measure of its jump component, where $\Upsilon$ satisfies $\int\min\{1,y^2\}\Upsilon(dy)<\infty$ and has negative support. The Laplace exponent $\psi$ of $\{Y\}$ fully characterizes its distributions, through its characteristic function $u\in\mathbb{R}\mapsto\E\left[\exp\left(\mathrm{i}u Y(t)\right)\right]=\exp\left[\psi(\mathrm{i}u)t\right]$.

Equation (\ref{eq:levykitch}) gives the most common parameterization of $\psi$. It corresponds to the L\'{e}vy-It\^{o} decomposition of $\{Y\}$ in a Brownian motion with linear drift $\tilde\mu t$, a compound Poisson process with jumps in $(-\infty,-1]$, and a pure-jump martingale with jumps in $(-1,0)$  \citep[][Section I.1]{cup96:bertoin}. Alternative parameterizations arise if we decompose the jumps of $\{Y\}$ in small and large shocks in other ways. These parameterizations all have the same dispersion parameter $\sigma$ and L\'{e}vy measure $\Upsilon$, but have different drift parameters. For example, in the special case that $\int_{(-1,0)}y\Upsilon(dy)<\infty$, the {\em compensator} term for the small shocks in (\ref{eq:levykitch}),
$\int_{(-\infty,0)}s y I(y>-1)\Upsilon(d y)=s\int_{(-1,0)}y \Upsilon(d y)$,
is a well-defined linear function of $s$. Therefore, in this case, we can alternatively parameterize $\psi$ as
\begin{equation}
\label{eq:levykitchBV}
\psi(s)=\mu s+\frac{\sigma^2}{2}
s^2+\int_{(-\infty,0)}\left(\mathrm{e}^{s y}-1\right)\Upsilon(d y),
\end{equation}

\noindent where $\mu\equiv\tilde\mu+\int_{(-1,0)}y \Upsilon(d y)$. This includes the important special case that $\int_{(-\infty,0)}\Upsilon(dy)<\infty$, in which $\{Y\}$ is the sum of a Brownian motion with drift parameter $\mu$ and a compound Poisson process with jumps of sizes in $(-\infty,0)$. In general, any of the equivalent parameterizations of $\psi$ can be used in the MHT model's specification, but some are numerically and statistically more convenient than others; we return to this in Section \ref{ss:parameterization}.

With $\psi$ determined, we are ready to analyze the Laplace transform $\LT(\cdot|x,v)$. The Laplace exponent, as a function on $\Rp$, is continuous and convex, and satisfies $\psi(0)=0$ and, because $\{Y\}$ is not weakly decreasing,
$\lim_{s\rightarrow\infty}\psi(s)=\infty$. Therefore, there exists a largest solution $\Lambda(0)\geq 0$ to $\psi(\Lambda(0))=0$ and an
inverse $\Lambda:[0,\infty)\rightarrow[\Lambda(0),\infty)$ of the restriction of $\psi$ to $[\Lambda(0),\infty)$. Theorem 1 of
\citet[Chapter VII]{cup96:bertoin} implies that $\LT(s|x,v)=\exp\left[-\Lambda(s)\phi(x)v\right]$ \citep[][Section 4.1]{ecta12:abbring}.
Using iterated expectations, the Laplace transform $\LT(\cdot|x)$ of the distribution $F(\cdot|x)$ of $T|X=x$ follows from
\begin{equation}
\label{eq:LaplaceTX}
\begin{split}
\LT(s|x)	&=\int_{\Rp}\exp\left(-s t\right)dF(t|x)=\int_{\Rpp}\left[\int_{\Rp}\exp\left(-s t\right)dF(t|x,v)\right]dG(v)\\
	&=\int_{\Rpp}\exp\left[-\Lambda(s)\phi(x)v\right]dG(v)=\L\left[\Lambda(s)\phi(x)\right],
\end{split}
\end{equation}
with $\L$  the Laplace transform of the distribution $G$ of $V$.

\subsection{Nontrivial Gaussian Component}
\label{ss:modelBM}

To facilitate the numerical computation of the MHT model's likelihood and ensure standard conditions for the maximum likelihood estimator, we assume throughout the paper's remainder that $\{Y\}$ has a nontrivial Gaussian component:
\begin{assumption}[{\bf Nontrivial Gaussian Component}] 
\label{ass:sigma}
$\psi$ satisfies (\ref{eq:levykitch}) with $\sigma>0$.
\end{assumption}

Assumption \ref{ass:sigma} excludes the case that $\{Y\}$ is a pure-jump process. To motivate this assumption, first consider the special case that $\{Y\}$ itself is a nontrivial Brownian motion, i.e. a Brownian motion with general drift coefficient $\mu\in\R$ and dispersion coefficient $\sigma\in\Rpp$ (obviously, this case satisfies Assumption \ref{ass:sigma}). Then, $\psi(s)$ equals  $\psi_{\mathrm{BM}}(s;\mu,\sigma)\equiv \mu s+\sigma^2 s^2/2$, so that
$\Lambda(0)$ equals $\Lambda_{\mathrm{BM}}(0;\mu,\sigma)\equiv\min\{0,-2\mu/\sigma^2\}$ and $\Lambda(s)$ equals
\begin{equation}
\Lambda_{\mathrm{BM}}(s;\mu,\sigma)\equiv \frac{\sqrt{\mu^2+2\sigma^2 s}-\mu}{\sigma^2}.
\label{eq:bminvform}
\end{equation}

\noindent For later reference, we have made the dependence on the parameters $\mu$ and $\sigma$ explicit here. Because there are no jumps, there is no ambiguity in the treatment of small and large jumps, and this parameterization of $\psi$ is unique. In particular, the L\'{e}vy-Khintchine representations (\ref{eq:levykitch}) and (\ref{eq:levykitchBV}) of $\psi$ coincide, and $\mu=\tilde\mu$.

In this special case, the distribution of $T|X=x,V=v$ is known to be inverse Gaussian, with explicit expressions for its Lebesgue density and survival function (see Section \ref {ss:MLBM}). If $\mu\geq 0$, then $\Lambda_{\mathrm{BM}}(0;\mu,\sigma)=0$ and the distribution of $T|X=x,V=v$ is nondefective.
If $\mu<0$, however, $\Lambda_{\mathrm{BM}}(0;\mu,\sigma)=-2\mu/\sigma^2>0$ and the
distribution of $T|X=x,V=v$ has a defect of size $1-\exp(2\phi(x)v\mu/\sigma^2)$. 
Either way, the MHT model specifies a mixed inverse Gaussian distribution for $T|X=x$ in this special case.\footnote{Mixed inverse Gaussian distributions have been used to model duration data in the statistical literature. For example, \citet{ss01:aalengjessing} proposed such a model with parametric mixing over the Brownian motion's drift coefficient $\mu$.}   Because this distribution has a Lebesgue density with full (and thus parameter-independent) support, it is straightforward to specify the likelihood for a parametric specification of $\phi$ and $G$ and to compute the corresponding maximum likelihood estimator, and this estimator will have standard asymptotic properties.

If $\{Y\}$ is a more general spectrally-negative L\'{e}vy process, then $F(\cdot|x)$ may have parameter-dependent support. For example, if $Y(t)=\mu t$, then $T(\phi(x)v)=\mu^{-1}\phi(x)v$, so that $F(\cdot|x)$ is concentrated on the support of $\mu^{-1}\phi(x)V$.  Assumption \ref{ass:sigma} excludes this pathology.
\begin{lemma}[{\bf Absolute Continuity}]
\label{lemma:ac}
If Assumption \ref{ass:sigma} holds then, for given $(x,v)\in{\cal X}\times\Rpp$ and some positive density $f(\cdot|x,v)$, $F(t|x,v)=\int_0^tf(u|x,v)du$ for all $t\in\Rp$. 
\end{lemma}
\begin{proof}
Because $\phi(x)v>0$ and $\lim_{s\rightarrow\infty}\Lambda(s)=\infty$, $F(0|x,v)=\lim_{s\rightarrow\infty}{\cal F}(s|x,v)=\lim_{s\rightarrow\infty}\exp\left[-\Lambda(s)\phi(x)v\right]=0$. Moreover, by Assumption \ref{ass:sigma}, for given $t\in\Rpp$, the distribution of $Y(t)$ is the convolution of a normal distribution and the distribution of the cumulated jumps, and therefore has a positive Lebesgue density on $\mathbb{R}$. Using that and $\phi(x)v>0$, \citet[][Chapter VII, Corollary 3]{cup96:bertoin} implies that $F(\cdot|x,v)$ has a positive Lebesgue density $f(\cdot|x,v)$ on $\Rpp$, and $F(t|x,v)=\int_0^tf(u|x,v)du$ for all $t\in\Rp$. 
\end{proof}

\noindent Note that, by Lemma \ref{lemma:ac} and Fubini's theorem, Assumption \ref{ass:sigma} also implies that $F(t|x)=\int_0^tf(u|x)du$, for all $t\in\Rp$, with positive Lebesgue density  $f(\cdot|x)\equiv\int_0^\infty f(\cdot|x,v) dG(v)$. Thus, Assumption \ref{ass:sigma} ensures that a standard parametric maximum likelihood approach can be used, as in the purely Gaussian case. A complication is that the distribution $F(\cdot|x)$ and its density $f(\cdot|x)$ are generally not known in closed form and need to be computed by inverting their Laplace transforms.  As we will see in Section \ref{ss:MLgeneral}, Assumption \ref{ass:sigma} facilitates a crucial computational simplification of this inversion. Moreover, in the next section, we will see that Assumption \ref{ass:sigma},  together with \cites{ecta12:abbring} assumptions and innocuous normalizations, suffices for the model's point identification.

\subsection{Nonparametric Identification}
\label{ss:identification}

The MHT model's primitives are $\psi$, $\phi$, and $G$. By \citet[][Section XIII.1, Theorem 1]{wiley71:fellerII}, there is a one-to-one relation between a probability distribution and its Laplace transform. Thus, we can equivalently write the primitives as $\psi$, $\phi$, and $\L$. By (\ref{eq:LaplaceTX}) and the definition of $\Lambda$, each specification of such an MHT triplet $(\psi,\phi,\L)$ implies a Laplace transform $\LT(\cdot|x)$ of the distribution $F(\cdot|x)$, and thus $F(\cdot|x)$ itself, for all $x\in{\cal X}$. 

One may wonder whether, conversely, knowledge of ${\cal F}(\cdot|x)$, $x\in{\cal X}$, would allow one to uniquely determine (``identify'') the model's primitives $(\psi,\phi,\L)$, perhaps after imposing some normalizations and restrictions. To be practical, we explicitly take into account that data on $T$ and $X$ will not allow us to determine ${\cal F}(\cdot|x)$ if $\Pr(X=x)=0$. So, suppose that we can determine ${\cal F}(\cdot|X)$ up to almost sure equivalence; that is, that we know $\mathbb{E}\left[{\cal F}(\cdot|X)I(X\in B)\right]=\mathbb{E}\left[\exp\left(-sT\right)I(X\in B)\right]$ for all measurable $B\subseteq{\cal X}$. Section \ref{ss:sampling} assumes a simple type of independent right censoring scheme for which this is true: random sampling from $(\min\{T,C\},I(T\leq C),X)$, with $T$ and $X$ drawn from the joint distribution of $(T,X)$ implied by some marginal distribution of $X$ and the model's conditional distribution $F(\cdot|x)$, $x\in{\cal X}$, and, for given $X$, the censoring time $C$ drawn, independently from $T$, from a conditional distribution such that $\Pr(C\geq t|X)>0$ for all $t\in\Rp$.\footnote{From the censored data, both the subdensity $f(t|X)\Pr(C\geq t|X)$, for almost all $t$, and the joint survival function $\Pr(T\geq t,C\geq t|X)=\overline F(t|X)\Pr(C\geq t|X)$ are identified up to almost sure equivalence. Thus, the hazard rate $f(t|X)/\overline F(t|X)= f(t|X)\Pr(C\geq t|X)/\Pr(T\geq t,C\geq t|X)$ is identified for almost all $t$, which determines ${\cal F}(\cdot|X)$, up to almost sure equivalence. See e.g. \citet{meth62:cox}. This argument extends to more general forms of independent censoring \citep[see e.g.][]{springer93:andersenetal}.}  Note that this includes the case in which we have ``complete'' observations from the joint distribution of $(T,X)$ (if $C=\infty$ always) and extends to more general independent censoring schemes.

Following \citet[][Section 3]{aos01:gillrobins}, we deal with the ambiguity arising from conditioning on (possibly) continuous covariates by assuming continuity of their effects. Let $B(x,\epsin)$ be an open ball of radius $\epsin>0$ around $x\in\mathbb{R}^K$. The support ${\cal X}$ of $X$ contains all  points $x\in{\cal X}$ such that $\Pr(X\in B\left(x,\epsin)\right)>0$ for all $\epsin>0$.

\begin{assumption}[{\bf Continuity of the Covariate Effects}]
\label{ass:cont}
The function $\phi$ and support ${\cal X}$ of $X$ are such that, for each $x\in{\cal X}$, $\lim_{\epsin\downarrow 0}\sup_{x'\in B(x,\epsin)\cap {\cal X}}|\phi(x')-\phi(x)|=0$.
\end{assumption}

\noindent For isolated mass points $x\in{\cal X}$, $B(x,\epsin)\cap{\cal X}=\{x\}$ for small enough $\epsin$, and Assumption \ref{ass:cont} does not constrain $\phi$. For points $x$ such that $B(x,\epsin)\subseteq{\cal X}$ for some $\epsin>0$, Assumption \ref{ass:cont} simply requires continuity of $\phi$, as a function on $\mathbb{R}^K$, at $x$. If $X$ has both finitely discrete and continuous components, then Assumption \ref{ass:cont} requires continuity of $\phi$ in the continuous components for given values of the discrete components. Assumption \ref{ass:cont} is satisfied if, for example, $\phi(x)=\exp(x'\beta)$ for some parameter vector $\beta\in\mathbb{R}^K$.

\begin{lemma}[{\bf Identification of the Conditional Distribution}]
\label{lemma:calF}
If Assumption \ref{ass:cont} holds, then 
\begin{equation}
\label{eq:calF}
{\cal F}(s|x)=\lim_{\epsin\downarrow 0}\frac{\mathbb{E}\left[\exp(-sT)I(X\in B(x,\epsin))\right]}{\mathbb{E}\left[I(X\in B(x,\epsin))\right]},~~~s\in\Rp,~x\in{\cal X}.
\end{equation}
\end{lemma}
\begin{proof}
By Assumption \ref{ass:cont} and continuity of ${\cal G}$, for every $\epsout>0$, there exists a $\epsin>0$ such that
$|{\cal F}(s|x')-{\cal F}(s|x)|=|{\cal G}\left(\Lambda(s)\phi(x')\right)-{\cal G}\left(\Lambda(s)\phi(x)\right)|<\epsilon$ for all $x'\in B(x,\epsin)$, so that
$\left|{\cal F}(s|x)-\mathbb{E}\left[\exp(-s T)I(X\in B(x,\epsin))\right]/\mathbb{E}\left[I(X\in B(x,\epsin))\right]\right|
<\epsout$.
\end{proof}

\noindent Note that, if $x$ is an isolated point in ${\cal X}$, then \eqref{eq:calF}  reduces to ${\cal F}(s|x)=\mathbb{E}\left[\exp(-sT)|X=x\right]$.

Following \citet{ecta12:abbring}, our identification analysis exploits variation of the threshold with the covariates.
\begin{assumption}[{\bf Nontrival Covariate Effects}]
\label{ass:variation}
For some $x_0,x_1\in{\cal X}$, $\phi(x_0)\neq\phi(x_1)$.
\end{assumption}
\noindent As is clear from the proof of the following theorem, under Assumption \ref{ass:cont}, the covariate values $x_0$ and $x_1$ in Assumption \ref{ass:variation} can be identified with values such that $F(\cdot|x_0)\neq F(\cdot|x_1)$.
\begin{theorem}[{\bf Nonparametric Identification}]
\label{th:identifiability} Let $(\psi,\phi,\L)$ and $(\tilde{\psi},\tilde{\phi},\widetilde{\L})$ be MHT
triplets that satisfy Assumptions \ref{ass:sigma}--\ref{ass:variation} and are observationally equivalent (imply the same conditional distribution $F(\cdot|X)$ up to almost sure equivalence). Then, for some $a,b\in\Rpp$: $\tilde\psi(s)=\psi(a s)$ and $\widetilde{\L}(s)=\L(b s)$ for all $s\in \Rp$, and $\tilde\phi=a b^{-1}\phi$.
\end{theorem}
\begin{proof}
By Assumption \ref{ass:cont} and Lemma \ref{lemma:calF}, we can identify ${\cal F}(\cdot|x)$ for all $x\in{\cal X}$. In particular, we can identify $x_0,x_1\in{\cal X}$ such that ${\cal F}(s|x_0)=\L\left[\Lambda(s)\phi(x_0)\right] \neq \L\left[\Lambda(s)\phi(x_1)\right]={\cal F}(\cdot|x_1)$, which exist by Assumption \ref{ass:variation}. Take these $x_0$ and $x_1$ as given.

We have that $(\psi;\phi(x_0),\phi(x_1);\L)$ and $(\tilde{\psi};\tilde{\phi}(x_0),\tilde{\phi}(x_1);\widetilde{\L})$ imply the same identified ${\cal F}(\cdot|x_0)$ and ${\cal F}(\cdot|x_1)$, and that ${\cal F}(\cdot|x_0)\neq {\cal F}(\cdot|x_1)$. This is the two-sample problem studied by \citet{ecta12:abbring}. We first apply \citeauthor{ecta12:abbring}'s Theorem 1, with Assumption \ref{ass:sigma}, to this two-sample problem and then extend the argument to the full domain ${\cal X}$ of $\phi$ and $\tilde\phi$. 

The L\'{e}vy-Khintchine formula (\ref{eq:levykitch}),  $\int\min\{1,y^2\}\Upsilon(dy)<\infty$, and dominated convergence imply that
$
\psi'(s)=\tilde\mu+\sigma^2
s+\int_{(-\infty,0)}\left\{y\mathrm{e}^{s y}-y
I(y>-1)\right\}\Upsilon(d y)$.
Using dominated convergence once more, it follows that $\lim_{s\rightarrow\infty}s^{-1}\psi'(s)=\sigma^2$. With Assumption \ref{ass:sigma}, this gives
$
\lim_{s\rightarrow\infty}\psi'(w s)/\psi'(s)=\lim_{s\rightarrow\infty}w(w s)^{-1}\psi'(w s)/\left[s^{-1}\psi'(s)\right]=w
$  
for all $w\in\Rpp$. The same is true for $\tilde\psi'$. Thus, both $|\psi'|$ and $|\tilde\psi'|$ vary regularly with exponent $1$ at infinity \citep[][Section VIII.8]{wiley71:fellerII}. Consequently, \citet[][Theorem 1]{ecta12:abbring} applies with $\rho=1$. Noting that \citeauthor{ecta12:abbring}'s setup, unlike ours, imposes a scale normalization on $\phi$, this implies that, for some $a,b\in\Rpp$, $\tilde\Lambda=a^{-1}\Lambda$ and $\widetilde{\L}(s)=\L(b s)$ for all $s\in\Rp$. The inverse of $\Lambda$ equals the restriction of $\psi$ to $[\Lambda(0),\infty)$ and can be uniquely analytically extended to its full domain $\Rp$; the same is true for the inverse of $\tilde\Lambda$. This gives $\tilde\psi(s)=\psi(a s)$ for all $s\in \Rp$.

Finally, fix any $s\in\Rpp$. Because ${\cal F}(\cdot|x)$ is identified, observational equivalence implies that $\L\left[\Lambda(s)\phi(x)\right]={\cal F}(s|x)=\widetilde{\L}\left[\tilde{\Lambda}(s)\tilde\phi(x)\right]=\L\left[\Lambda(s)a^{-1}b\tilde\phi(x)\right]$ for all $x\in{\cal X}$. Therefore, $\tilde\phi= a b^{-1}\phi$.
\end{proof}

\noindent The first part of the proof, which establishes the relation between $(\psi,\L)$ and $(\tilde\psi,\widetilde{\L})$, only uses Assumption \ref{ass:cont} for continuity at $x_0$ and $x_1$. So, we can relax Assumption \ref{ass:cont} accordingly if we weaken Theorem \ref{th:identifiability}'s claim that $\tilde\phi=a b^{-1}\phi$ to $\tilde\phi(X)=a b^{-1}\phi(X)$ almost surely. 

Unlike the model studied by \citet{ecta12:abbring}, our model with a nontrivial Gaussian component is identified, up to two unknown scale parameters $a$ and $b$. It is easy to see why $a$ and $b$ cannot be determined by data on $T$ and $X$ alone. Mixed hitting times $T(\phi(X)V)$ are not affected by rescaling both the latent process $\{Y\}$ and the threshold $\phi(X)V$ by the same factor, nor by rescaling the threshold factors $\phi(X)$ and $V$ without changing the threshold itself. Specifically, suppose that $(\psi,\phi,\L)$ in Theorem \ref{th:identifiability} corresponds to a latent process $\{Y\}$ and threshold $\phi(X)V$. Then, the observationally equivalent $(\tilde{\psi},\tilde{\phi},\widetilde{\L})$ corresponds to a latent process $\{a Y\}$, an observed threshold factor $a b^{-1} \phi(X)$, and an unobserved threshold factor $b V$. Clearly, the implied first hitting times are the same:
$\inf\left\{t\geq 0:Y(t)>\phi(X)V\right\}=\inf\left\{t\geq 0:a Y(t)>a b^{-1}\phi(X)b V\right\}$.
Identification therefore requires that the scales of two of $\{Y\}$, $\phi(X)$ and $V$ are normalized. The most convenient way of implementing these normalizations depends on the chosen parameterization.

\subsection{Parameterization and Normalization}
\label{ss:parameterization}

This paper's estimation procedure requires a computationally feasible, flexible parameterization of the model. To this end, we specify the L\'{e}vy measure $\Upsilon(\cdot;\alpha)$ up to a finite vector of unknown parameters $\alpha$. With a drift parameter $\mu$ and Gaussian dispersion parameter $\sigma$,  this specification and the L\'{e}vy-Khintchine formula (in our proposed specifications, (\ref{eq:levykitchBV})) imply a parameterization $\psi(\cdot;\mu,\sigma,\alpha)$ of the Laplace exponent. We similarly specify $\phi(\cdot;\beta)$, and $\L(\cdot;\kappa)$ up to finite vectors $\beta$ and $\kappa$ and collect all parameters in $\theta\equiv(\mu,\sigma,\alpha,\beta,\kappa)$. We make sure that the proposed parameterizations are unique, in the sense that different values of $\theta$ map into different primitives $\psi(\cdot;\mu,\sigma,\alpha)$, $\phi(\cdot;\beta)$, and $\L(\cdot;\kappa)$. We also discuss ways to normalize them. A corollary to Theorem \ref{th:identifiability} then establishes parametric identification.

\paragraph{Latent process}
Recall that $\Upsilon(\cdot;\alpha)=0$ and the Laplace exponent equals $\psi_{\mathrm{BM}}(s;\mu,\sigma)=\mu s+\frac{\sigma^2}{2} s^2$, with $\sigma>0$, if $\{Y\}$ is a nontrivial Brownian motion with drift. We distinguish this basic specification with a subscript ``BM'' because it appears in our computations for more general specifications of $\psi(\cdot;\alpha)$ as well. We consider two such specifications.  

The first adds an independent compound Poisson process with a finitely discrete shock distribution to the basic specification. Because $\int_{(-1,0)}y\Upsilon(dy;\alpha)<\infty$ in this case, the L\'{e}vy-Khintchine formula (\ref{eq:levykitchBV}) now offers the simplest way to parameterize $\psi$:
$\psi(s;\mu,\sigma,\alpha)=\mu
s+\frac{\sigma^2}{2} s^2+\sum_{j=1}^J \lambda_j\left(\mathrm{e}^{s
\nu_j}-1\right)$,
where $\alpha\equiv(\lambda_1,\ldots,\lambda_J,\nu_1,\ldots,\nu_J)$, with 
$\lambda_j>0$ the Poisson rate at which shocks of size $\nu_j<0$ arrive; $j=1,\ldots,J$; and $\nu_1<\ldots<\nu_J$.\footnote{Equivalently, in this specification, shocks
arrive at a rate $\lambda\equiv\sum_{j=1}^J\lambda_j$ and are drawn
independently from a distribution with $J$ points of support
$(\nu_1,\ldots,\nu_J)$ with probabilities
$\left(\lambda_1/\lambda,\ldots,\lambda_J/\lambda\right)$. 
We exclude the boundary cases in which $\lambda_j=0$, $\nu_j=0$, or $\nu_{j-1}=\nu_j$, which correspond to specifications with fewer than $J$ shock sizes, to ensure a unique parameterization and standard inference. See Footnote \ref{fn:discrete}.} 

The second specification  instead assumes that shocks arrive at a Poisson rate $\lambda$ and have sizes drawn from a gamma distribution with density
$\frac{\omega^\tau}{\Gamma(\tau)} \: (-y)^{\tau-1}\exp(\omega y)$; $\omega,\tau>0$; at $y\in(-\infty,0)$.  We can again use (\ref{eq:levykitchBV}), which now gives $\psi(s;\mu,\sigma,\alpha)=\mu s+\frac{\sigma^2}{2}s^2+\lambda\left\{(s/\omega+1)^{-\tau}-1\right\}$, where $\alpha\equiv(\lambda,\omega,\tau)$.

The L\'{e}vy-Khintchine formula (\ref{eq:levykitchBV}) provides a unique parameterization of the Laplace exponent in terms of the drift parameter $\mu$, the Gaussian dispersion parameter $\sigma$, and the L\'{e}vy measure $\Upsilon$.\footnote{\citet[][Chapter 1, Theorem 1]{cup96:bertoin} and the discussion following it show that the general L\'{e}vy-Khintchine formula (\ref{eq:levykitch}) provides a unique parameterization of the Laplace exponent in terms of $\tilde\mu$, $\sigma$, and $\Upsilon$. Consequently, formula (\ref{eq:levykitchBV}) does as well with, as discussed in Section \ref{ss:char}, a different drift parameter.} In turn, our two specifications of the jump process give unique parameterizations of $\Upsilon$. Consequently, both parameterizations $\psi(\cdot;\alpha)$ are unique. 

The scale of $\psi(\cdot;\mu,\sigma,\alpha)$ can be normalized by setting $|\mu|=1$, which implicitly assumes that $\mu\neq 0$, or $\sigma=1$. After all, if $\psi(\cdot;\mu,\sigma,\alpha)$ is a Laplace exponent with $|\mu|=1$ (or $\sigma=1$) then, for $a>0$, $s\mapsto\psi(a s;\mu,\sigma,\alpha)$ is a Laplace exponent with $|\mu|=a$ (or $\sigma=a$).\footnote{One can alternatively normalize the scale of the jump component, which varies across specifications.}

\paragraph{Covariate effects}
The threshold is naturally specified to be loglinear in the covariates: $\phi(x;\beta)=\exp(x'\beta)$. Note that this specification implies Assumption \ref{ass:cont}. 

Suppose that ${\cal X}\subseteq\mathbb{R}^K$ is not contained in a proper linear subspace of $\mathbb{R}^K$. Then, this parameterization is unique: $\exp(x'\tilde\beta)=\exp(x'\beta)$ for all $x\in{\cal X}$ implies that $\beta=\tilde\beta$. Moreover, it embodies a scale normalization: For given $\beta$ and $a\in\Rpp/\{1\}$,  there exists no $\tilde\beta$ such that $a\phi(x;\alpha)=\exp(\ln(a)+x'\beta)=\exp(x'\tilde\beta)$.

\paragraph{Unobserved heterogeneity}
We entertain a finitely discrete specification of $G$. This specification is versatile, computationally convenient, and appears naturally in \cites{ecta84:heckmansinger} work on semi-nonparametric estimation of the MPH model.  It assumes that $V$ has $L\in\mathbb{N}$ support points $0<v_1<\cdots<v_L$, with $0<\pi_l\equiv\Pr(V=v_l)<1$; $l=1,\ldots,L$. Then, $\L(s;\kappa)=\sum_{l=1}^L\pi_l\exp(-s v_l)$, with $\kappa\equiv(v_1,\ldots,v_L,\pi_1,\ldots,\pi_{L-1})$ and $\pi_L\equiv 1-\sum_{l=1}^{L-1}\pi_l$.\footnote{\label{fn:discrete}We assume that all $\pi_l\in(0,1)$ and that all support points are distinct to ensure that the parameterization of $G$ is unique. In practice, we may want to include the boundary cases, because these correspond to specifications with fewer than $L$ support points. This, however, leads to nonstandard identification and inference, because we can either reduce the number of support points from $L$ to $L-1$ by setting $\pi_L=0$, in which case $v_L$ is irrelevant, or by setting $v_{L-1}=v_L$, in which case only $\pi_{L-1}+\pi_L$ matters.} The inequality constraints ensure that the parameterization is unique. It  can be scale normalized by setting $v_1=1$.

   
\begin{corollary}[{\bf Parametric Identification}] 
\label{corr:parsamident}
Let $\theta$ and $\tilde\theta$, via one of this section's parameterizations, map into observationally equivalent MHT triplets. Suppose that Assumptions \ref{ass:sigma} and \ref{ass:variation} hold, ${\cal X}\subseteq\mathbb{R}^K$ is not contained in a proper linear subspace of $\mathbb{R}^K$, and either $\phi$ or $\L$ is scale normalized. Then, $\theta=\tilde\theta$. 
\end{corollary}

\noindent Corollary \ref{corr:parsamident} does not rely on the fact that the finitely discrete specification of $G$ ensures that $\mathbb{E}[V]<\infty$, which would suffice for identification without Assumption 1 \citep[see][Section 4.3]{ecta12:abbring}. We maintain  Assumption \ref{ass:sigma}, because it is essential to our approach to estimation (see Section \ref{ss:modelBM}) and allows for alternative specifications of $G$ that do not imply $\mathbb{E}[V]<\infty$. This may, for example, be useful in an extension to sieve estimation, in which it may be hard to impose $\mathbb{E}[V]<\infty$  (see Section \ref{s:concl}).


\section{Maximum Likelihood Estimation}
\label{s:ML}

Fix one of the previous section's parameterizations $\theta\mapsto[\psi(\cdot;\mu,\sigma,\alpha),\phi(\cdot;\beta),\L(\cdot;\kappa)]$. Denote the implied parametric density  of $T|X=x$ with $f(\cdot|x;\theta)$ and the corresponding survival function with $\overline F(\cdot|x;\theta)$. Similarly, write $f(\cdot|x,v;\theta)$ and $\overline F(\cdot|x,v;\theta)$. This section presents a method for evaluating this parameterization's likelihood for a basic but common sampling scheme, using the Gaussian special case as a benchmark. 

\subsection{Sampling and Likelihood}
\label{ss:sampling}

Let $\left\{(T_1,X_1),\ldots,(T_N,X_N)\right\}$ be a random sample from the distribution of $(T,X)$ induced by $F(\cdot|x;\theta_0)$, $x\in{\cal X}$, at the ``true'' parameter vector $\theta_0$ and some marginal distribution of $X$. We do not directly observe this complete sample, but only a censored version of it: $\left\{(T_1^*,D_1,X_1)\ldots,(T_N^*,D_N,X_N)\right\}$. Here, $T_n^*\equiv\min\{T_n,C_n\}$ is the observed duration and $D_n\equiv I(T_n\leq C_n)$  a censoring indicator, for some random censoring time $C_n$. Note that a complete observation $(T_n^*,D_n)=(t,1)$ pairs an MHT event $T_n=t$ with a censoring event $C_n\geq t$, whereas a censored observation $(T_n^*,D_n)=(t,0)$ corresponds to $T_n>t$ and $C_n=t$. 

We assume a simple type of independent right-censoring \citep{springer93:andersenetal}. Suppose that $(T_n,C_n,X_n)$ is independent across $n$ and that, conditional on $X_n$, $C_n$ is independent of $T_n$, with a distribution that does not depend on $\theta_0$. Then, conditional on $X_n$, the likelihood contribution of $(T_n^*,D_n)$ factorizes in an MHT part, $f(T_n^*|X_n;\theta)^{D_n}{\overline F}(T_n^*|X_n;\theta)^{1-D_n}$, and a censoring part that does not depend on $\theta$. Thus, the conditional likelihood is proportional to $\prod_{n=1}^Nf(T_n^*|X_n;\theta)^{D_n}{\overline F}(T_n^*|X_n;\theta)^{1-D_n}$. Its maximizer is the full-information maximum likelihood estimator of $\theta_0$ if the covariates $X_n$ carry no information on $\theta_0$. 

Note that the case without censoring, so that $T^*_n=T_n$ and $D_n=1$ almost surely for all $n$, is included as a special case in which $C_n=\infty$ almost surely for all $n$. Also, with more general independent right censoring schemes, the resulting estimator remains a valid (but often, partial) likelihood estimator \citep{springer93:andersenetal}. Moreover, the likelihood, and the corresponding estimator, can easily be adapted to other practically relevant sampling schemes, such as those involving interval censoring.

\subsection{Gaussian Special Case}
\label{ss:MLBM}

Suppose that $\{Y\}$ is a Brownian motion with drift, so that, by the analysis in Section \ref{ss:modelBM}, $T|X$ has a mixed inverse Gaussian distribution. Then, up to a constant containing the censoring time events, the log conditional (on the covariates) likelihood $\ell_N(\theta )$  equals
\begin{equation}
\label{eq:GaussianLikelihood}
\ell_N(\theta)=\sum_{n=1}^N \ln\int f_{\mathrm{BM}}(T^*_n|X_n,v;\mu,\sigma,\beta)^{D_n}\overline F_{\mathrm{BM}}(T^*_n|X_n,v;\mu,\sigma,\beta)^{1-D_n}dG(v;\kappa),
\end{equation}

\noindent where
\begin{equation}
\label{eq:inverseGaussianpdf}
f_{\mathrm{BM}}(t|x,v;\mu,\sigma,\beta)=\frac{\phi(x;\beta)v}{\sigma\sqrt{2\pi t^3}}\exp\left(-\frac{[\phi(x;\beta)v-\mu t]^2}{2\sigma^2 t}\right)
\end{equation}

\noindent is the Lebesgue density of the inverse Gaussian distribution and
\begin{equation}
\label{eq:inverseGaussianFbar}
\overline{F}_{\mathrm{BM}}(t|x,v;\mu,\sigma,\beta) 
=\Phi\left(\frac{\phi(x;\beta)v-\mu t}{\sigma\sqrt{t}}\right)-\exp\left(\frac{2\mu \phi(x;\beta)v}{\sigma^2}\right)\Phi\left(-\frac{\phi(x;\beta)v+\mu t}{\sigma\sqrt{t}}\right)
\end{equation}

\noindent is its survival function \citep[][Section 5.4]{methuen65:coxmiller}. Here, $\Phi$ is the cumulative standard normal distribution function.  With Section \ref{ss:parameterization}'s finite discrete specification of $G$, the log likelihood in (\ref{eq:GaussianLikelihood}) reduces to
\begin{equation}
\label{eq:likDiscrete}
\ell_N(\theta)=\sum_{n=1}^N \ln \sum_{l=1}^L \pi_l f_{\mathrm{BM}}(T^*_n|X_n,v_l;\mu,\sigma,\beta)^{D_n}\overline F_{\mathrm{BM}}(T^*_n|X_n,v_l;\mu,\sigma,\beta)^{1-D_n}.
\end{equation}
If we e.g. specify $\phi(x;\beta)=\exp(x'\beta)$, this log likelihood, its derivatives, and  its maximizer $\hat\theta_N$ are easy to compute using (\ref{eq:inverseGaussianpdf}) and (\ref{eq:inverseGaussianFbar}). Under standard regularity conditions, including the normalizations and assumptions needed for Corollary \ref{corr:parsamident}'s parametric identification, $\hat\theta_N$ is a consistent and asymptotically normal estimator of $\theta_0$. Given the assumption that the marginal distribution of $X$ and the censoring times carry no information on $\theta_0$, it is also asymptotically efficient. Its asymptotic covariance matrix can quickly be estimated using either the score or Hessian characterization of the Fisher information matrix. 

Many of the models studied in the statistics literature similarly lead to explicit expressions for the likelihood that facilitate estimation \citep{ss06:leewhitmore}. In the general L\'{e}vy case, such explicit expressions are not available, and maximum likelihood cannot be implemented directly. The next section develops methods for computing the maximum likelihood estimator and its asymptotic distribution in this general case.

\subsection{General Case}
\label{ss:MLgeneral}

In general, $f(\cdot|x;\theta)$ and ${\overline F}(\cdot|x;\theta)$ are not explicitly known, but can be computed by numerically inverting their Laplace transforms. Our approach is based on the work of \citet{japr00:rogers}, who applied a variant of \cites{abate:92} inversion method to the problem of calculating the first-passage-time distribution of a spectrally one-sided L\'{e}vy process. 

Following \citeauthor{japr00:rogers}, we first consider calculating the survival function ${\overline F}(\cdot|x;\theta)$. Using integration by parts,
it is easy to show that its Laplace transform ${\overline\LT}(s|x;\theta)\equiv\int_{0}^{\infty} \exp(-st) \overline{F}(t|x;\theta) d t=s^{-1}\left\{1-\LT\left(s|X\right)\right\}$.  So, for given $\theta$, we can explicitly construct ${\overline\LT}(s|x;\theta)=s^{-1}\left\{1-\L\left[\Lambda(s;\mu,\sigma,\alpha)\phi(x;\beta);\kappa\right]\right\}$ and obtain $\overline{F}(\cdot|x;\theta)$ using \textit{Mellin's inverse formula} \citep[e.g.][]{davies:02},
\begin{equation}
\overline{F}(t|x;\theta) = \frac{1}{2\pi\mathrm{i}}\lim_{\xi\rightarrow\infty}\int_{\gamma_\xi}\exp(st){\overline\LT}(s|x;\theta)d s.
\label{eq:mellin}
\end{equation}

\noindent Here, the integration is along the contour $\gamma_\xi:u\in[-1,1]\mapsto\pathconst+\mathrm{i}\xi u$, which traces out a straight line  in $\mathbb{C}$, parallel to the imaginary axis from $\pathconst-\mathrm{i}\xi$ to $\pathconst+\mathrm{i}\xi$. We make this contour's dependence on $\pathconst\in\mathbb{R}$ explicit by writing $\gamma_\xi(u;\pathconst)$ for its value at $u$. The parameter $\pathconst$ should be chosen such that it is larger than the real part of any singularity in the Laplace transform ${\overline \LT}(\cdot|x;\theta)$. Because ${\overline \LT}(\cdot|x;\theta)$ is analytic on the set of all $s$ with $\Re\,s>0$, we can choose any $\pathconst>0$.

The integral in (\ref{eq:mellin}) does not generally have an explicit solution, but can be efficiently approximated using numerical methods. A key complication is that our specification of ${\overline \LT}(\cdot|x;\theta)$ involves the inverse function $\Lambda$, which cannot generally be expressed in closed form. To circumvent this problem, we follow \citeauthor{japr00:rogers} and instead integrate along the composition $\tilde\gamma_\xi\equiv\psi\circ\Lambda_{\mathrm{BM}}\circ\gamma_\xi$, which is a contour in $\mathbb{C}$ from $\psi\left[\Lambda_{\mathrm{BM}}\left(\pathconst-\mathrm{i}\xi; \mu,\sigma\right); \mu,\sigma,\alpha\right]$ to $\psi\left[\Lambda_{\mathrm{BM}}\left(\pathconst+\mathrm{i}\xi; \mu,\sigma\right); \mu,\sigma,\alpha\right]$. Here,  $\Lambda_{\mathrm{BM}}$ is the inverse of the Laplace exponent of the Brownian motion component of $\psi$, for which (\ref{eq:bminvform}) gives an explicit expression. Note that $\Lambda_{\mathrm{BM}}$ necessarily has the same dispersion parameter $\sigma$ as $\psi$, but that its drift parameter is not uniquely pinned down (because the drift parameter of $\psi$ depends on the way we deal with small shocks; see Section \ref{ss:char}). Fortunately, the exact value of the drift parameter of $\Lambda_{\mathrm{BM}}$ plays no role in the argument that follows. It can generally be set to the drift parameter in the specific parameterization of $\psi$ used; for example, $\tilde\mu$ in (\ref{eq:levykitch}) or $\mu$ in (\ref{eq:levykitchBV}). Following Section \ref{ss:parameterization}'s specifications of $\psi$ with compound Poisson jumps, we have set the drift parameter of $\Lambda_{\mathrm{BM}}$ equal to $\mu$ in (\ref{eq:levykitchBV}). We make the transformed contour's dependence on $\pathconst$ and the parameters of $\psi$ explicit by writing $\tilde\gamma_\xi(u;\mu,\sigma,\alpha,\pathconst)$ for its value at $u$. 

\citeauthor{japr00:rogers} argued that, under Assumption \ref{ass:sigma}, replacing $\gamma_\xi$ by $\tilde\gamma_\xi$ in (\ref{eq:mellin}) does not affect that integral's value, so that
\begin{equation}
\overline{F}(t|x;\theta) = \frac{1}{2\pi\mathrm{i}}\lim_{\xi\rightarrow\infty}\int_{\tilde\gamma_\xi}\exp(st) \overline{\LT}(s;x;\theta) d s
=\frac{1}{2\pi\mathrm{i}}\lim_{\xi\rightarrow\infty}\int_{\gamma_\xi}\overline{q}^*(t,s|x;\theta)ds,
\label{eq:newinv}
\end{equation}
with 
\[\begin{split}
&\overline{q}^*(t,s|x;\theta)\equiv\\
&~\exp\left\{\psi\left[\Lambda_{\mathrm{BM}}(s;\mu,\sigma)\right]t;\mu,\sigma,\alpha\right\}\frac{1-\L\left[\Lambda_{\mathrm{BM}}(s;\mu,\sigma)\phi(x;\beta);\kappa\right]}{\psi\left[\Lambda_{\mathrm{BM}}(s;\mu,\sigma)\right]} \frac{d}{d s}\psi\left[\Lambda_{\mathrm{BM}}(s;\mu,\sigma);\mu,\sigma,\alpha\right];
\end{split}\]

\noindent which no longer involves $\Lambda$. This argument relies on Cauchy's integral theorem, which implies that an integral  over the analytic integrand in (\ref{eq:mellin}) along a closed contour equals zero. This is particularly true for the closed contour formed by going up $\gamma_\xi$ from $\gamma_\xi(-1;\pathconst)$ to $\gamma_\xi(1;\pathconst)$, crossing over from $\gamma_\xi(1;\pathconst)$ to $\tilde\gamma_\xi(1;\mu,\sigma,\alpha,\pathconst)$, going down  $\tilde \gamma_\xi$ from $\tilde\gamma_\xi(1;\mu,\sigma,\alpha,\pathconst)$ to $\tilde\gamma_\xi(-1;\mu,\sigma,\alpha,\pathconst)$, and crossing back from $\tilde\gamma_\xi(-1;\mu,\sigma,\alpha,\pathconst)$ to $\gamma_\xi(-1;\pathconst)$. Consequently, the integrals in (\ref{eq:mellin}) and (\ref{eq:newinv}) are equal, provided that the integrals over the contour from $\gamma_\xi(1;\pathconst)$ to $\tilde\gamma_\xi(1;\mu,\sigma,\alpha,\pathconst)$ and the contour from $\gamma_\xi(-1;\pathconst)$ to $\tilde\gamma_\xi(-1;\mu,\sigma,\alpha,\pathconst)$ vanish as $\xi\rightarrow\infty$. \citeauthor{japr00:rogers} concluded that this is the case, because the integrand vanishes sufficiently fast along these two contours as $\xi\rightarrow\infty$ (in particular,  $s{\overline \LT}(s|x;\theta)\rightarrow 1$ as $|s|\rightarrow\infty$) and, under Assumption \ref{ass:sigma}, their lengths do not grow too fast with $\xi$. In particular, 
\[
\begin{split}
\left|\frac{\gamma_\xi(1;\pathconst)-\tilde\gamma_\xi(1;\mu,\sigma,\alpha,\pathconst)}{\gamma_\xi(1;\pathconst)}\right|
&=\left|\frac{\pathconst+\mathrm{i}\xi-\psi\left[\Lambda_{\mathrm{BM}}\left(\pathconst+\mathrm{i}\xi;\mu,\sigma\right);\mu,\sigma,\alpha\right]}{\pathconst+\mathrm{i}\xi}\right|\\
&=\left|\frac{\psi_{\mathrm{BM}}\left[\Lambda_{\mathrm{BM}}\left(\pathconst+\mathrm{i}\xi;\mu,\sigma\right);\mu,\sigma\right]-\psi\left[\Lambda_{\mathrm{BM}}\left(\pathconst+\mathrm{i}\xi;\mu,\sigma\right);\mu,\sigma,\alpha\right]}
{\psi_{\mathrm{BM}}\left[\Lambda_{\mathrm{BM}}\left(\pathconst+\mathrm{i}\xi;\mu,\sigma\right);\mu,\sigma\right]}\right|
\end{split}
\]

\noindent converges to zero as $\xi\rightarrow\infty$ (note that the right hand side of (\ref{eq:levykitch}) is dominated by the Gaussian term for large $s$). Similarly,
$\left|\frac{\gamma_\xi(-1;\pathconst)-\tilde\gamma_\xi(-1;\mu,\sigma,\alpha,\pathconst)}{\gamma_\xi(-1;\pathconst)}\right|\rightarrow 0$ as $\xi\rightarrow \infty$.

Using a change of variables, we can rewrite  (\ref{eq:newinv}) as an integral over the real line:
\begin{equation}
\overline{F}(t|x;\theta)=\frac{1}{2\pi}\int_{-\infty}^\infty \overline{\realint}(t,u|x;\theta,\pathconst)du,
\label{eq:newinv2}
\end{equation}
where $\overline{\realint}(t,u|x;\theta,\pathconst)\equiv \overline{q}^*(t,\pathconst+\mathrm{i}u|x;\theta)$. Following \citeauthor{abate:92}, we can apply the trapezoidal rule to approximate (\ref{eq:newinv2}) with the infinite sum
\begin{equation}
\overline{S}_{\infty}(t|x;\theta,\pathconst,h)  \equiv \frac{h}{2\pi}\sum_{r=-\infty}^{\infty} \Re\,\overline{\realint}(t,r h|x;\theta,\pathconst),
\label{eq:invsum}
\end{equation}
where $h>0$  is the rule's step size. Note that we only need to approximate the real part of (\ref{eq:newinv2}), because its imaginary part should be zero. \citeauthor{abate:92} discussed the error introduced by this discretization and noted that it works particularly well because the integrand oscillates and the approximation errors tend to cancel out. 

In practice, we need to truncate the infinite sum $\overline{S}_{\infty}(t|x;\theta,\pathconst,h)$ in (\ref{eq:invsum}) to 
$\overline{S}_{R}(t|x;\theta,\pathconst,h)\equiv \frac{h}{2\pi}\sum_{r=-R}^{R}\Re\,\overline{\realint}(t,r h|x;\theta,\pathconst)$
for some $R\in\mathbb{N}$ and use extrapolation to approximate the case where $R\rightarrow\infty$. Because $\overline{S}_{R}(t|x;\theta,\pathconst,h)$ is nearly periodic in $R$, $\lim_{R\rightarrow\infty} \overline{S}_{R}(t|x;\theta,\pathconst,h)$ can be efficiently approximated using Euler summation:
\begin{equation}
\overline{F}(t|x) \approx \overline{E}_{R,M}(t|x;\theta,\pathconst,h) \equiv \sum_{m=0}^{M} 2^{-M}\binom{M}{m}\overline{S}_{R+m}(t|x;\theta,\pathconst,h),
\label{eq:finapp}
\end{equation}

\noindent for some $M\in\mathbb{N}$. \citeauthor{abate:92} proposed to estimate the associated error by $\overline{E}_{R,M+1}(t|x;\theta,\pathconst,h)-\overline{E}_{R,M}(t|x;\theta,\pathconst,h)$. In our case, this estimate quickly tends to zero as M is increases, which suggests that the approximation is accurate (see also Section \ref{s:experiments}).

We follow a similar procedure to calculate the density $f(\cdot|x;\theta)$ from its Laplace transform $\LT(\cdot|x;\theta)$. We again start with Mellin's inverse formula (\ref{eq:mellin}) with contour $\gamma_\xi$, but now with $f(t|x;\theta)$ in its left hand side and $\LT(s|x;\theta)$ in its right hand side. With the finitely discrete specification of $G$,  $\LT(s|x;\theta)$ vanishes more rapidly than $\overline{\LT}(s|x;\theta)$ ($s\LT(s|x;\theta)\rightarrow 0$, whereas $s{\overline \LT}(s|x;\theta)\rightarrow 1$) as $|s|\rightarrow\infty$.\footnote{This follows from the fact  that the behavior of $\LT(s|x;\theta)$ for large $s$ is dominated by the term $\pi_1\exp\left\{-\Lambda(s;\mu,\sigma)\phi(x;\beta)v_1\right\}$ corresponding to the lowest support point $v_1$ of $G$. With specifications of $G$ that have support near zero, $\LT(s|x;\theta)$ may vanish more slowly than $\overline{\LT}(s|x;\theta)$ as $|s|\rightarrow\infty$. For example, if $G$ is a gamma distribution, one can show that $|s{\overline \LT}(s|x;\theta)|\rightarrow\infty$ as $|s|\rightarrow\infty$. Simulations suggest our procedure is nevertheless accurate in this case.\label{fn:gamma}} This suggests that we can again replace the contour $\gamma_\xi$ in Mellin's inverse formula with $\tilde\gamma_\xi$ and that 
\[ 
f(t|x;\theta)=\frac{1}{2\pi\mathrm{i}}\lim_{\xi\rightarrow\infty}\int_{\gamma_\xi}q^*(t,s|x;\theta)ds,
\]
where
\[\begin{split}
&q^*(t,s|x;\theta)\equiv\\
&~\exp\left\{\psi\left[\Lambda_{\mathrm{BM}}(s;\mu,\sigma)\right] t;\mu,\sigma,\alpha\right\}\L\left[\Lambda_{\mathrm{BM}}(s;\mu,\sigma)\phi(x;\beta);\kappa\right] \frac{d}{ds}\psi\left[\Lambda_{\mathrm{BM}}(s;\mu,\sigma);\mu,\sigma,\alpha\right].
\end{split}\]
As before, we can rewrite this into an integral over the real line,
\[
f(t|x;\theta)=\frac{1}{2\pi}\int_{-\infty}^\infty \realint(t,u|x;\theta,\pathconst)du,
\]
where $\realint(t,u|x;\theta,\pathconst)\equiv q^*(t,\pathconst+\mathrm{i}u|x;\theta)$, and approximate this integral with an Euler sum $E_{R,M}(t|x;\theta,\pathconst,h)$. 

One could control the computation of $f(t|x;\theta)$ and $\overline{F}(t|x;\theta)$ with different tuning parameters $c$, $h$, $R$, and $M$. However, as our notation $E_{R,M}(t|x;\theta,\pathconst,h)$ and $\overline{E}_{R,M}(t|x;\theta,\pathconst,h)$ for the corresponding Euler sums suggests, we will not do so in this paper. We take guidance from \citeauthor{japr00:rogers} in setting the common values of $c$, $h$, $R$, and $M$. In the next sections, we find that his suggestion to use duration-$t$ specific values $\pathconst=11/t$ and $h=\pi/t$ yields good numerical performance in our case. We will adopt these as our default settings, together with $R=9$ and $M=25$.\footnote{\citet{japr00:rogers} claimed that $R=6$ and $M=15$ trade off accuracy and speed well. Because of the advances in computing speed since then, we can opt for more accuracy. See Section \ref{s:experiments} for some details.} 

The log likelihood for an independently censored sample satisfies 
\begin{equation}
\begin{split}
\label{eq:likApprox}
\ell_N(\theta)&=\sum_{n=1}^N D_n\ln f(T^*_n|X_n;\theta)+(1-D_n)\ln\overline{F}(T^*_n|X_n;\theta)\\
&\approx\sum_{n=1}^N D_n\ln E_{R,M}(T^*_n|X_n;\theta,\pathconst,h)+(1-D_n)\ln\overline{E}_{R,M}(T^*_n|X_n;\theta,\pathconst,h).
\end{split}
\end{equation}

\noindent We have implemented an estimator in MATLAB that maximizes this approximate log likelihood using a quasi-Newton algorithm with BFGS updates for the Hessian and multiple random starting values \citep{nocedal:06}. 

We supply an analytical gradient of the approximate log likelihood with respect to the parameter vector $\theta$ to ensure quick and stable maximization. This gradient sums contributions of the $N$ observations. Consider the contribution of observation $n$. Suppose that this observation is complete ($D_n=1$; the calculations for a censored observation are similar). The approximate likelihood contribution of this observation, $E_{R,M}(T^*_n|X_n;\theta,\pathconst,h)$, is the real part of a weighted sum of $\realint(T^*_n,r h|X_n;\theta,\pathconst)$ over finitely many values of $r$, with weights that do not depend on $\theta$. Each term $\realint(T^*_n,r h|X_n;\theta,\pathconst)$ in this weighted sum is the product of three factors; 
\[
\exp\left[\psi\left(z;\mu,\sigma,\alpha\right) T^*_n\right], ~~~ \L\left[z\phi(X_n;\beta);\kappa\right], ~~~ \text{and} ~~ \psi'\left(z;\mu,\sigma,\alpha\right)\Lambda'_{\mathrm{BM}}\left(c+\mathrm{i}rh;\mu,\sigma\right);
\] 
that are smooth in $\theta$ and $z$, composed with $z=\Lambda_{\mathrm{BM}}(c+\mathrm{i}rh;\mu,\sigma)$, which is itself smooth in $\mu$ and $\sigma$. Its complex-valued derivative with respect to $\theta$ follows from tedious but straightforward application of the product and chain rules. We ignore the imaginary part of the weighted sum of these derivatives over $r$, because the imaginary part of the likelihood contribution $f(T^*_n|X_n;\theta)$ that we approximate with $E_{R,M}(T^*_n|X_n;\theta,\pathconst,h)$ is zero. So, we set the contribution of observation $n$ to the gradient of the log likelihood equal to the real part of this weighted sum of derivatives, divided by $E_{R,M}(T^*_n|X_n;\theta,\pathconst,h)$. The analytical gradient sums these contributions. We construct asymptotic standard errors from the corresponding Hessian, which we calculate using finite differences of the analytical gradient. The replication package \citep{zenodo21:abbringsalimans} provides further details.

The MATLAB code currently normalizes $\psi(\cdot|\mu,\sigma,\alpha)$ by setting $\mu=1$. Note that this implicitly assumes that $\mu>0$. It would be straightforward to adapt the code to instead normalize $|\mu|=1$, which more generally allows for $\mu\neq 0$, or $\sigma=1$, which does not restrict $\mu$ at all. 

Our estimator maximizes an approximate log likelihood. For some applications, it has been shown that the maximum approximate likelihood estimator is first order equivalent to the exact maximum likelihood estimator if the approximations improve sufficiently quickly with the sample size \citep[e.g.][]{ecta02:aitsahalia}. We could try to derive a similar equivalence result for our estimator, using \citeauthor{abate:92}'s numerical analysis and some further results on the tail behavior of $\overline{\realint}(t,u|x;\theta,\pathconst)$ and ${\realint}(t,u|x;\theta,\pathconst)$. However, as we will see in Section \ref{s:experiments}, we can compute our estimator very accurately  in reasonable time, so that a formal result establishing how accuracy should increase with sample size would not be of much practical use. Therefore, we take the pragmatic approach that much of the literature has taken and simply apply standard maximum likelihood asymptotics.\footnote{This is how \citet{singleton2001estimation} handled his maximum likelihood estimator of a discretely sampled affine diffusion, which, like our estimator, required numerical Fourier inversion. He expressed some worries about the computational burden of his Fourier inversion procedure, but only for the multivariate case. We only use univariate Fourier inversion and benefit from 20 years of computational development.}


\section{Numerical Experiments}
\label{s:experiments}

We have investigated the accuracy of the proposed likelihood approximation by conducting a range of numerical experiments. We discuss the results of three of these experiments here. All three experiments use the default settings for the parameters that control the approximation, unless explicitly stated otherwise. The first two experiments directly compare the explicitly known duration density and likelihood implied by MHT models without shocks to their approximations. The third experiment focuses on a model with shocks, for which the implied duration density is not known in explicit form.

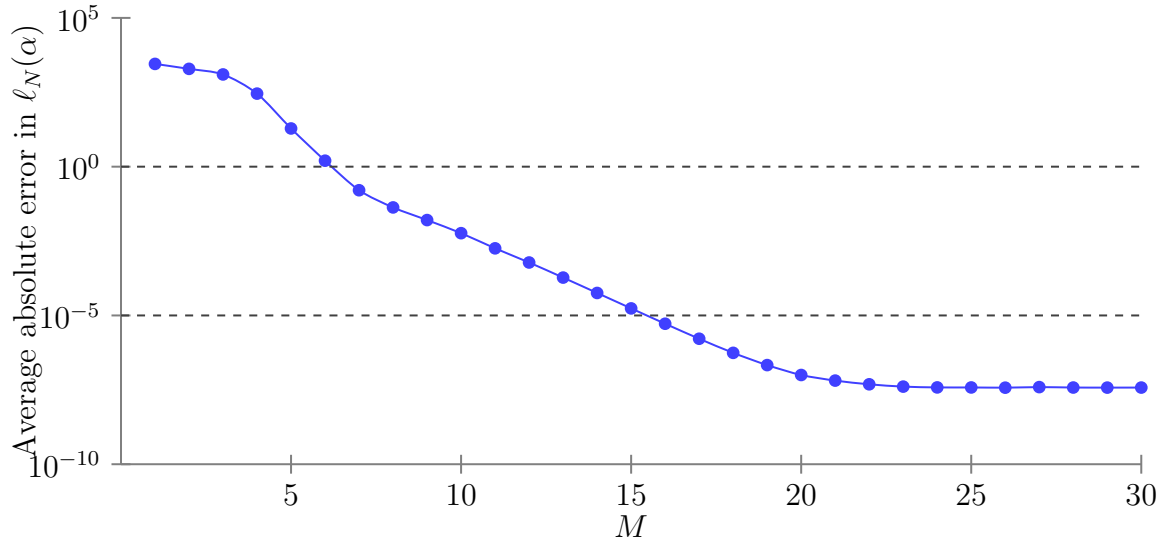
\begin{figure}[t]
\caption{Approximation Error of the Log Likelihood for Various $M$\label{fig:llherr}}
\medskip
\begin{centering}
\begin{tikzpicture}
\tikzstyle{every axis y label}=[yshift=10pt]
\tikzstyle{every axis x label}=[yshift=-5pt]
   \begin{axis}[width=15cm,height=7.5cm,
        name=llherr,
        ylabel={Average absolute error in $\ell_N(\alpha)$},
        xlabel={$M$},
        axis x line=bottom,
        every outer x axis line/.append style={-,color=gray,line width=0.75pt},
        axis y line=left,
        every outer y axis line/.append style={-,color=gray,line width=0.75pt},
        xtick={5,10,15,20,25,30},
        xticklabels={$5$,$10$,$15$,$20$,$25$,$30$},
        ytick={-10,-5,0,5},
        yticklabels={$10^{-10}$,$10^{-5}$,$10^0$,$10^5$},
        major tick length=2.5mm,
        every tick/.append style={line width=0.75pt},
        ymin=-10,
        ymax=5,
       xmin=0,
       xmax=30,
        scaled ticks=false,
        /pgf/number format/precision=2,
        /pgf/number format/set thousands separator={}]
        \addplot[color=blue!75,smooth,mark=*,line width=0.75pt] table[x=M,y=llherr,col sep=comma]{mht-likelihood/fig1.csv} ;
         \draw[color=black!75,line width=0.75pt,dashed] (axis cs:0,-5) -- (axis cs:30,-5);
         \draw[color=black!75,line width=0.75pt,dashed] (axis cs:0,0) -- (axis cs:30,0);
\end{axis}
\end{tikzpicture}
\end{centering}

\noindent {\footnotesize Note: This figure is based on the log likelihood $\ell_N(\theta)$ of an MHT model with a Brownian motion latent process and discrete unobserved heterogeneity with four support points for \cites{jem85:kennan} complete strike duration data. It plots the average absolute difference between $\ell_N(\theta)$ and its numerical approximation over 100 randomly drawn parameter values $\theta$, for a range of values of $M$. The errors are plotted on a logarithmic scale. Throughout, $\mu$ and $\sigma^2$ are set equal to their maximum likelihood estimates for a simple inverse Gaussian model with $\phi(X;\beta)V=1$, which are known in closed form, and $v_1=1$. The remaining support points $v_2$, $v_3$, and $v_4$ of the heterogeneity distribution are generated by exponentiating draws from a standard normal distribution, so that they vary in level, but are all approximately of the right scale. All four support points $v_{l}$ receive probability mass $1/4$. The parameter $\beta$ multiplying the covariates is set to zero.}
\end{figure}

The first experiment compares direct computations of the log likelihood function of the mixed inverse Gaussian model using the explicit expression for the density in (\ref{eq:inverseGaussianpdf}) to its numerical approximations as we vary $M$. The log likelihood is calculated on the data set that we use in Section \ref{s:strike}. This ensures that this experiment  provides both a real life test case and a check on the results we present in that section. The data contain 566 complete strike durations. Because the approximation errors are close to unbiased, the error in the log likelihood scales with the root of the sample size.

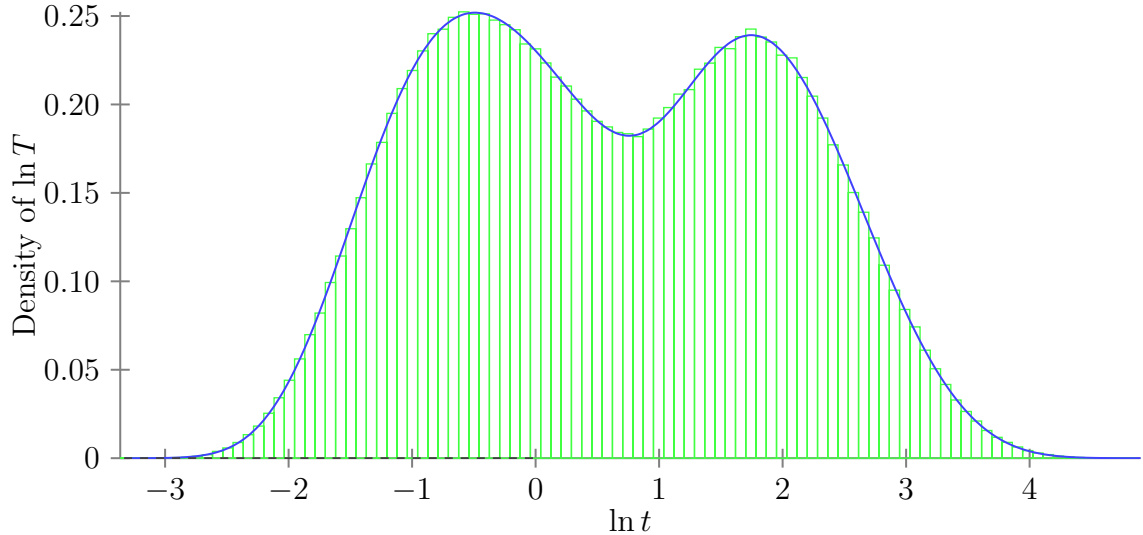
\begin{figure}[t]
\caption{Approximate Probability Density and Histogram of Simulated Values of $\ln T$ for a Specification With Shocks and Heterogeneity\label{fig:histogram}\label{fig:approxshock}}
\medskip
\begin{centering}
\begin{tikzpicture}
\tikzstyle{every axis y label}=[yshift=10pt]
\tikzstyle{every axis x label}=[yshift=-5pt]
   \begin{axis}[width=15cm,height=7.5cm,
        name=llherr,
        ylabel={Density of $\ln T$},
        xlabel={$\ln t$},
        axis x line=bottom,
        every outer x axis line/.append style={-,color=gray,line width=0.75pt},
        axis y line=left,
        every outer y axis line/.append style={-,color=gray,line width=0.75pt},
        ytick={0,0.05,0.10,0.15,0.20,0.25},
        yticklabels={$0$,$0.05$,$0.10$,$0.15$,$0.20$,$0.25$},
        major tick length=2.5mm,
        every tick/.append style={line width=0.75pt},
        scaled ticks=false,
        /pgf/number format/precision=2,
        /pgf/number format/set thousands separator={}]
        \addplot[color=green!75,ybar interval,line width=0.50pt] table[x=logtlow,y=hist,col sep=comma]{mht-likelihood/fig3hist.csv} ;
        \addplot[color=blue!75,smooth,line width=0.75pt] table[x=logt,y=invlt,col sep=comma]{mht-likelihood/fig3invlap.csv} ;
         \draw[color=black!75,line width=0.75pt,dashed] (axis cs:-28,0) -- (axis cs:0,0);
         \draw[color=black!75,line width=0.75pt,dashed] (axis cs:-28,-5) -- (axis cs:0,-5);
         \draw[color=black!75,line width=0.75pt,dashed] (axis cs:-28,-10) -- (axis cs:0,-10);
   \end{axis}
\end{tikzpicture}
\end{centering}

\noindent {\footnotesize Note: This figure plots the approximate probability density of $\ln T$ (smooth line) and a histogram of $1,000,000$ simulated values of $\ln T$ (bars), for an MHT model in which $\{Y\}$ equals a standard Brownian motion minus an independent compound Poisson process with mean $1/2$ exponential jumps at a rate of one per time unit ($\mu=\sigma=\tau=1$ and $\omega=2$) and the threshold equals $\phi(X;\beta)V=1$ with probability $0.7$ and $\phi(X;\beta)V=5$ with probability $0.3$.}
\end{figure}

Figure \ref{fig:llherr} plots the average of the absolute approximation error of the log likelihood, for different values of $M$, over 100 model parameters randomly generated at the scale of their maximum likelihood estimates. We find that this average absolute error decreases exponentially with $M$; this result is robust across the various parameter values over which the plotted results are averaged.
Consistently with \citet{japr00:rogers}, we see that $M=15$ already provides a decent approximation for most practical purposes. However, because the time required for the calculations grows only linearly in $M$, we can increase $M$ to 25 at a very low computational cost and obtain a nearly thousandfold increase in precision (with most of the gain already obtained with $M=20$). Once $M\geq 25$, other factors, such as rounding errors, become important, and the approximation error levels off. We also find that, with $M=25$, increasing $R$ or decreasing the step size $h$ adds very little to the precision of the inversion. The numerical approximation of the log likelihood takes 9--11 times as long to calculate as the analytical expression. However, in absolute terms this is still very manageable. For example, it takes about a second to calculate the density for a specification with shocks on a regular laptop computer 100,000 times.\footnote{We used Figure \ref{fig:approxshock}'s specification and MATLAB 2020b on a MacBook Pro (2018, 15inch, 2.9GHz 6-Core Intel Core i9, 32 GB 2400 MHz DDR4) with macOS 10.15.7.}  Consistently with this, the log likelihood can be maximized, starting from multiple random parameter values for each maximization, in under half a minute for the model specifications that we consider in Section \ref{s:strike}. 

\begin{figure}[t]
\caption{Approximation Error of the Log Inverse Gaussian Density Function\label{fig:proberr}}
\medskip
\begin{centering}
\begin{tikzpicture}
\tikzstyle{every axis y label}=[yshift=10pt]
\tikzstyle{every axis x label}=[yshift=-5pt]
   \begin{axis}[width=15cm,height=7.5cm,
        name=llherr,
        ylabel={Absolute error in $\ln f_{\mathrm{BM}}(t|X)$},
        xlabel={$\ln f_{\mathrm{BM}}(t|X)$},
        axis x line=bottom,
        every outer x axis line/.append style={-,color=gray,line width=0.75pt},
        axis y line=left,
        every outer y axis line/.append style={-,color=gray,line width=0.75pt},
        xtick={-25,-20,-15,-10,-5,0},
        xticklabels={$-25$,$-20$,$-15$,$-10$,$-5$,$0$},
        ytick={-10,-5,0},
        yticklabels={$10^{-10}$,$10^{-5}$,$10^0$},
        major tick length=2.5mm,
        every tick/.append style={line width=0.75pt},
        ymin=-12,
        ymax=2,
       xmin=-28,
       xmax=0,
        scaled ticks=false,
        /pgf/number format/precision=2,
        /pgf/number format/set thousands separator={}]
        \addplot[color=blue!75,only marks,mark=*,mark size=1pt] table[x=lap,y=logerr,col sep=comma]{mht-likelihood/fig2.csv} ;
         \draw[color=black!75,line width=0.75pt,dashed] (axis cs:-28,0) -- (axis cs:0,0);
         \draw[color=black!75,line width=0.75pt,dashed] (axis cs:-28,-5) -- (axis cs:0,-5);
         \draw[color=black!75,line width=0.75pt,dashed] (axis cs:-28,-10) -- (axis cs:0,-10);
\end{axis}
\end{tikzpicture}
\end{centering}
\noindent {\footnotesize Note: This figure plots the absolute difference between the log inverse Gaussian density $\ln f_{\mathrm{BM}}(t|X;\theta)$ with parameters $\mu=\sigma^2=\phi(X;\beta)V=1$ and its numerical approximation, on a logarithmic scale, against $\ln f_{\mathrm{BM}}(t|X;\theta)$, for a range of times $t$.}
\end{figure}
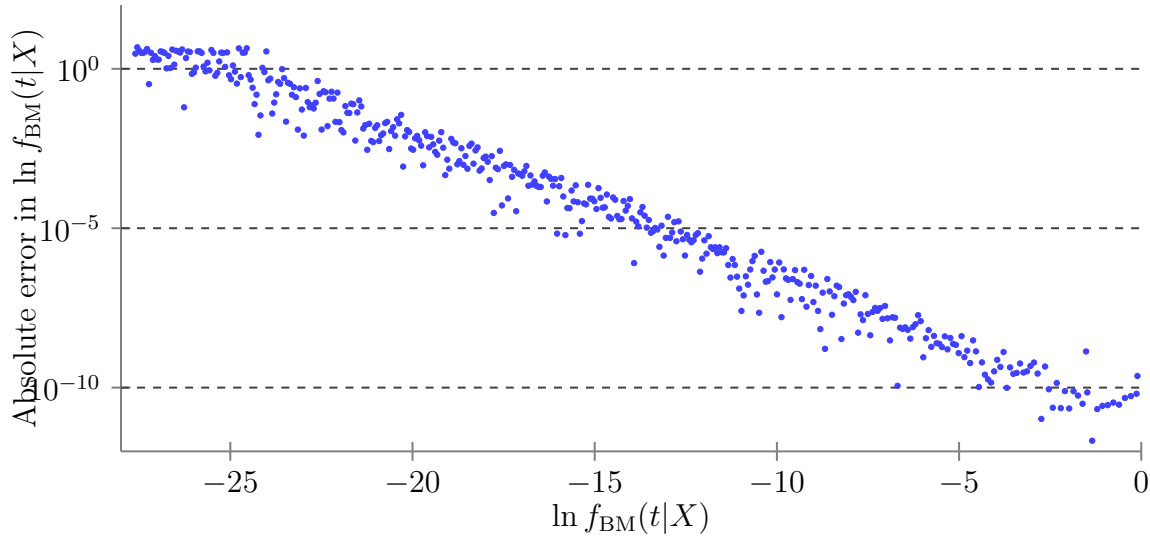

The second experiment takes a closer look at the numerical approximation of the density $f_{\mathrm{BM}}$ of a basic inverse Gaussian model with parameters such that $\mu=\sigma^{2}=\phi(X;\beta)V=1$. We only present results for $M=25$, but found very similar results for any $M \geq 20$. For the purpose of maximum likelihood estimation, we care most about the errors in the approximation of the {\em log} density, $\ln f_{\mathrm{BM}}$. Figure \ref{fig:proberr} plots the absolute error of this approximation against the log density itself, on a logarithmic scale. The (log-)linear relation displayed by the graph implies that the absolute error in the approximation of $\ln f_{\mathrm{BM}}(t|X;\theta)$ roughly equals $10^{-11}/f_{\mathrm{BM}}(t|X;\theta)$. Consequently, the approximation error is generally small, but the approximation breaks down when the density gets very small (say, $f_{\mathrm{BM}}(t|X;\theta)<10^{-10}$, or $\ln f_{\mathrm{BM}}(t|X;\theta)<-23$). When estimating the model with maximum likelihood, we can easily avoid this by setting reasonable starting values for the parameters. This ensures that the approximation is sufficiently precise for numerically robust maximum likelihood estimation.

The third experiment considers a model with shocks and a heterogeneous threshold. Figure \ref{fig:approxshock} plots the approximate density of $\ln T$ for this model, again using $M=25$. In this case, the true density is not explicitly known, so we compare the approximate density with a fine histogram of many simulated values of $\ln T$. Our approximate density closely tracks the simulated one. This finding is robust across model specifications.


\section{Strike Durations}
\label{s:strike}

The mere existence of nontrivial delays in labor agreements has puzzled economists; duration patterns in their
resolution have been studied to learn more about underlying bargaining games and information structures.

\citet{jrssa72:lancaster} analyzed strike durations using a Gaussian MHT model with regressors, but without unobserved heterogeneity. He interpreted the gap between the Brownian motion and the threshold as the level of
disagreement, and concluded that this model fits his data for the United Kingdom well. Others used proportional hazards models to study strike durations.  \cite{jem85:kennan}, in particular, showed that the US strike duration
hazard is $U$-shaped and took this as evidence against \citeauthor{jrssa72:lancaster}'s (homogeneous) MHT model.
He noted that this aspect of the data can be interpreted in terms of heterogeneity in the conflicts underlying the strikes, but did not subsequently pursue this in his empirical analysis.

Here, we will investigate whether \citeauthor{jem85:kennan}'s strike data can be matched well by a more general MHT model that explicitly takes into account unobserved heterogeneity in strikes. Such a model comes with \citeauthor{jrssa72:lancaster}'s attractive interpretation in terms of a level of disagreement that may both vary over time and initially be heterogeneous between strikes. We will explicitly discuss our estimation results in terms of this interpretation, with an implicit understanding that it is our modest objective to illustrate our methods and the descriptive and potential structural appeal of the MHT model, without providing a fully structural analysis of strike durations.

\cites{jem85:kennan} data cover all contract strikes in US manufacturing in the period 1968--1976 that involved at least a thousand workers, and that were classified to be primarily about ``general wage changes''. They include the durations in days of 566 strikes and, for each strike, a measure of the state of the business cycle in the month it started: the residuals of a regression of log industrial production in US manufacturing on linear and quadratic trend terms and seasonal dummies. We obtained the data in a fixed format text file {\tt strkdur.asc} from  \cites{cup05:camerontrivedi} web page. We divided all strike durations by seven, so that they are measured in weeks.

\def\vone{$   1.1$}
\def\vtwo{$   3.2$}
\def\vthree{$   7.2$}
\def\vfour{$  18.6$}
\begin{table}
\caption{Maximum Likelihood Estimates for \cites{jem85:kennan} Strike Duration Data\label{table:strike}}
\vspace*{0.5em}
\begin{center}
\small{\begin{tabular}{ccccccc}
\toprule
& I & II & III & IV & V & VI\tabularnewline
\midrule
\midrule
$\mu$ & $1$ & $1$ & $1$ & $1$ & $1$ & $1$ \tabularnewline
& $(0)$ & $(0)$ & $(0)$ & $(0)$ & $(0)$ & $(0)$ \tabularnewline
\midrule
$\sigma^{2}$ & $19.659$ & $ 6.218$ & $ 2.067$ & $ 1.227$ & $ 1.197$ & $ 0.542$\tabularnewline
& $( 3.157)$ & $( 0.863)$ & $( 0.403)$ & $( 0.217)$ & $( 0.218)$ & $( 0.315)$\tabularnewline
\midrule
$\lambda$ & $   $ & $   $ & $   $ & $   $ & $   $ & $ 0.019$\tabularnewline
& $$ & $$ & $$ & $$ & $$ & $( 0.021)$\tabularnewline
\midrule
$\nu$ & $   $ & $   $ & $   $ & $   $ & $   $ & $-5.133$\tabularnewline
& $$ & $$ & $$ & $$ & $$ & $( 2.546)$\tabularnewline
\midrule
$\beta$ & $-0.931$ & $-1.772$ & $-1.085$ & $-0.867$ & $-0.862$ & $-0.579$\tabularnewline
& $( 0.601)$ & $( 0.687)$ & $( 0.643)$ & $( 0.628)$ & $( 0.629)$ & $( 0.611)$\tabularnewline
\midrule
$v_1$ & $ 6.260$ & $ 2.543$ & $ 1.537$ & $ 1.105$ & $ 1.031$ & $ 0.755$\tabularnewline
& $( 0.467)$ & $( 0.199)$ & $( 0.142)$ & $( 0.113)$ & $( 0.175)$ & $( 0.177)$\tabularnewline
\midrule
$v_2$ & $   $ & $ 8.751$ & $ 5.888$ & $ 3.209$ & $ 1.756$ & $ 2.083$\tabularnewline
& $$ & $( 0.520)$ & $( 0.390)$ & $( 0.452)$ & $( 1.032)$ & $( 0.510)$\tabularnewline
\midrule
$v_3$ & $   $ & $   $ & $18.161$ & $ 7.165$ & $ 3.518$ & $ 4.138$\tabularnewline
& $$ & $$ & $( 1.011)$ & $( 0.560)$ & $( 0.763)$ & $( 0.842)$\tabularnewline
\midrule
$v_4$ & $   $ & $   $ & $   $ & $18.557$ & $ 7.303$ & $ 7.412$\tabularnewline
& $$ & $$ & $$ & $( 0.698)$ & $( 0.645)$ & $( 0.552)$\tabularnewline
\midrule
$v_5$ & $   $ & $   $ & $   $ & $   $ & $18.575$ & $17.004$\tabularnewline
& $$ & $$ & $$ & $$ & $( 0.693)$ & $( 1.220)$\tabularnewline
\midrule
$\pi_1$ & $     1$ & $ 0.399$ & $ 0.353$ & $ 0.252$ & $ 0.199$ & $ 0.198$\tabularnewline
& $(     0)$ & $( 0.044)$ & $( 0.034)$ & $( 0.038)$ & $( 0.117)$ & $( 0.040)$\tabularnewline
\midrule
$\pi_2$ & $   $ & $ 0.601$ & $ 0.492$ & $ 0.283$ & $ 0.098$ & $ 0.201$\tabularnewline
& $$ & $( 0.044)$ & $( 0.034)$ & $( 0.050)$ & $( 0.133)$ & $( 0.073)$\tabularnewline
\midrule
$\pi_3$ & $   $ & $   $ & $ 0.154$ & $ 0.315$ & $ 0.256$ & $ 0.223$\tabularnewline
& $$ & $$ & $( 0.023)$ & $( 0.053)$ & $( 0.083)$ & $( 0.062)$\tabularnewline
\midrule
$\pi_4$ & $   $ & $   $ & $   $ & $ 0.151$ & $ 0.297$ & $ 0.238$\tabularnewline
& $$ & $$ & $$ & $( 0.019)$ & $( 0.064)$ & $( 0.064)$\tabularnewline
\midrule
$\pi_5$ & $   $ & $   $ & $   $ & $   $ & $ 0.150$ & $ 0.140$\tabularnewline
& $$ & $$ & $$ & $$ & $( 0.019)$ & $( 0.020)$\tabularnewline
\midrule
\midrule
$\ell_N$ & $-1658.9$ & $-1588.7$ & $-1583.0$ & $-1576.3$ & $-1576.1$ & $-1575.4$\tabularnewline

\bottomrule
\end{tabular}}
\end{center}
{\footnotesize Note: The drift is normalized to $1$ per week. All specifications include a single covariate, \cites{jem85:kennan} deseasonalized and detrended log industrial production.  Asymptotic standard errors are in parentheses.}
\end{table}

Table \ref{table:strike} reports maximum likelihood estimates for a range of Section \ref{ss:parameterization}'s flexible parameterizations. All reported estimates are computed using Section \ref{ss:MLgeneral}'s numerical methods, with $M=25$. To further check these methods and their MATLAB implementation, we have also computed the same estimates for lower values of $M\geq 15$  (not reported), and estimates for the first five specifications using the explicit expressions for the log likelihood that are available in these cases (not reported). These results are virtually identical to those reported in Table \ref{table:strike}.

Columns I--V present estimates of models with Brownian motion latent processes and discrete unobserved heterogeneity. Throughout, the drift is normalized to 1 per week ($\mu=1$), so that $\E\left[T|X,V\right]=-\LT'(0+|X,V;\theta)=\exp(X'\beta)V$. By its construction as a regression residual, $X$ varies around zero and is close to zero on average  in the sample. Consequently, $V$ can be interpreted as the unobserved initial level of disagreement, measured as the mean number of strike weeks it commands.

The log likelihood substantially improves when adding a second, third and fourth support point to the distribution of $V$, between Columns I and IV, but a fifth support point (Column V) hardly changes the fit and the other parameters' estimates.  The estimates indicate that there is both substantial heterogeneity in the strikes' initial levels of disagreement and uncertainty in their evolution over time. The numbers in Column IV imply that there are four unobserved types of labor conflict, on average commanding respectively \vone, \vtwo, \vthree, and \vfour\ strike weeks. Each type's level of disagreement evolves with a standard deviation per week just above the unit drift towards agreement.

It is instructive to note that the variance of the latent process drops substantially, from close to 20 to just over 1, when more heterogeneity is added between Columns I and IV. Clearly, Column I's specification falsely attributes heterogeneity in the strikes'  initial levels of disagreement to uncertainty in their evolution over time.

The estimates of the coefficient $\beta$ reflect the effect of the business cycle on strike durations. In line with \cites{jem85:kennan} results,  strikes that begin in months with low production last longer. In the MHT model, this is captured by a countercyclical threshold: In times with low production, in expectation, conflicts command more strike days. One interpretation is that strike days are less costly in times with low production.  The precision of the estimates of $\beta$ is low. This is consistent with \citeauthor{jem85:kennan}'s results. He obtained more precise results with a binary cyclical indicator constructed from the indicator used here. For simplicity, we do not follow this lead here.

Column VI reports an estimate of a specification that includes discrete shocks of size $\nu$ at Poisson times. The estimates point to an infrequent shock  that sets back just over five weeks of drift towards agreement. The shock only somewhat improves the likelihood; a specification without shock, such as those in Columns IV and V, seems to be sufficient.

Finally, a very similar result is found with a gamma shock at a Poisson time (not reported). With this specification, virtually the same estimate of the arrival rate of the shocks is obtained. Moreover, the estimated gamma shock distribution is close to degenerate at Column VI's estimate of the shock size ($\nu$). Specifically, the estimates of the shape ($\tau$) and scale ($\omega$) parameters of the gamma distribution are both very large, and their ratio equals Column VI's estimated shock size. As expected, the same log likelihood is found.

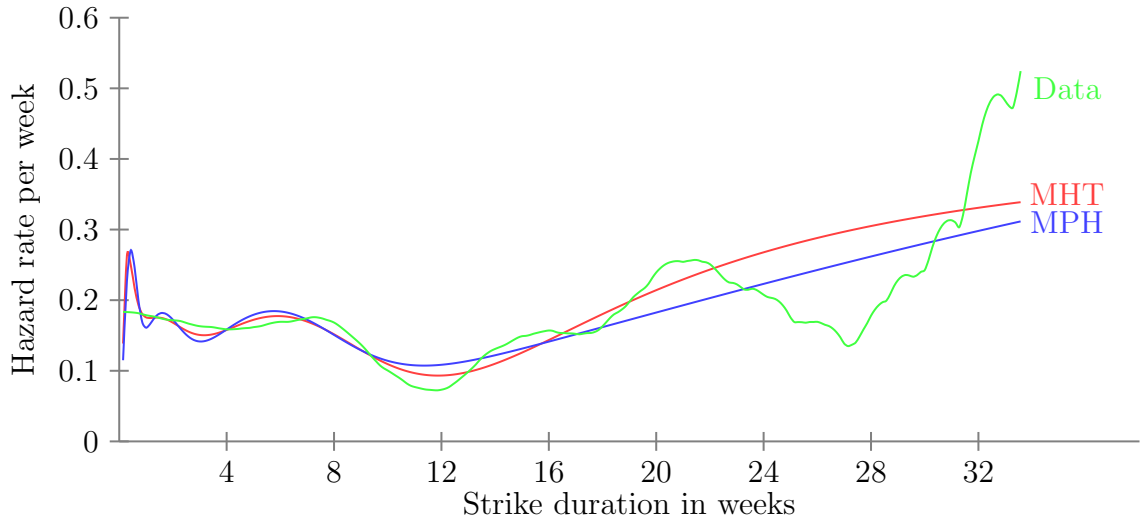
\begin{figure}[t]
\caption{Aggregate Strike End Hazard Rates\label{fig:strikehazard}}
\begin{centering}
\begin{tikzpicture}
\tikzstyle{every axis x label}=[yshift=-5pt]
   \begin{axis}[width=15cm,height=7.2cm,
        name=strikes,
        ylabel={Hazard rate per week},
        xlabel={Strike duration in weeks},
        axis x line=bottom,
        every outer x axis line/.append style={-,color=gray,line width=0.75pt},
        axis y line=left,
        every outer y axis line/.append style={-,color=gray,line width=0.75pt},
        xtick={4,8,12,16,20,24,28,32},
        xticklabels={$4$,$8$,$12$,$16$,$20$,$24$,$28$,$32$},
        ytick={0,0.1,0.2,0.3,0.4,0.5,0.6},
        yticklabels={$0$,$0.1$,$0.2$,$0.3$,$0.4$,$0.5$,$0.6$},
        major tick length=2.5mm,
        every tick/.append style={line width=0.75pt},
        ymin=0,
        ymax=0.6,
        xmin=0,
        xmax=38,
        scaled ticks=false,
        /pgf/number format/precision=2,
        /pgf/number format/set thousands separator={}]
        \addplot[color=red!75,smooth,line width=0.75pt] table[x=t,y=mht,col sep=comma]{mht-likelihood/fig4.csv} (axis cs: 35.3,0.35) node[color=red!75]{MHT};
        \addplot[color=blue!75,smooth,line width=0.75pt] table[x=t,y=mph,col sep=comma]{mht-likelihood/fig4.csv} (axis cs: 35.3,0.31) node[color=blue!75]{MPH};
        \addplot[color=green!75,smooth,line width=0.75pt] table[x=t,y=data,col sep=comma]{mht-likelihood/fig4.csv} (axis cs: 35.3,0.5) node[color=green!75]{Data};
\end{axis}
\end{tikzpicture}
\end{centering}
\vspace*{3mm}

\footnotesize{Note: This graph plots the empirical strike end hazard rate (Data), computed with Epanechnikov kernel smoothing from \cites{jem85:kennan} data, and the corresponding hazards implied by estimated MHT and MPH models. For the MHT model, the estimates in Table \ref{table:strike} for a specification with a latent Brownian motion and a discrete unobserved heterogeneity distribution with four support points are used. For the MPH model, we use maximum likelihood estimates of a model with the same discrete heterogeneity distribution and a Weibull baseline. Estimated hazard rates of the unconditional distribution of $T$ are plotted, based on the estimated distributions of $T|X$ implied by the models and the empirical distribution of the covariate $X$.}
\end{figure}

Figure \ref{fig:strikehazard} plots the aggregate hazard implied by the MHT model's estimates in Column IV of Table \ref{table:strike}. It also plots the hazard implied by estimates a MPH model with a Weibull baseline and a discrete heterogeneity distribution with four support points. Note that this MPH specification has exactly the same number of parameters as Column IV's MHT specification. In both cases, we computed the distribution of $T|X$ implied by these estimates, integrated over the empirical distribution of $X$, and computed and plotted the hazard rate of the resulting distribution. Figure \ref{fig:strikehazard}  also plots the empirical hazard rate, computed by kernel smoothing the raw data.

\def\mphllh{$-1577.9$}
\def\diffllh{$   1.6$}

Both the MHT and the MPH models fit the empirical hazard well, but the MPH model's log likelihood, at \mphllh, is \diffllh\ points lower. Because the Weibull baseline is monotonic, the Weibull MPH model can only fit the nonmonotonic strike hazard by compensating an increasing baseline hazard with negative duration dependence due to unobserved heterogeneity. Of course, usually MPH models with richer specifications of the baseline hazard are estimated and a sufficiently rich specification can fit the empirical hazard arbitrarily well.


\section{Conclusion}
\label{s:concl}

The results in this paper enable applied researchers to analyze duration data with mixed hitting-time (MHT) models using standard likelihood-based estimation and inference methods. The MATLAB code for parametric maximum likelihood estimation that accompanies this paper can directly be applied to either complete or independently right-censored duration data, and is easy to adapt to more general censoring schemes. 

Our procedure for likelihood computation lends itself well for use in semi-nonparametric maximum likelihood estimation \citep[e.g.][]{elsevier07:chen}. As in \cite{ecta84:heckmansinger}'s analysis of the MPH model, we could handle unobserved heterogeneity nonparametrically using discrete heterogeneity distributions with a varying number of support points.  Some care would have to be taken to ensure that the likelihood approximation continues to work well if the unobserved heterogeneity, in the limit, has support near zero (see Footnote \ref{fn:gamma}). Similarly, the L\'{e}vy-It\^{o} decomposition of $\{Y\}$ (see Section \ref{ss:char}) suggests that we construct a sieve for $\psi$ using Section \ref{ss:parameterization}'s specification that sums a Gaussian component with an independent compound Poisson component, with the shocks distributed discretely with a varying number of support points. This way, each element of the sieve satisfies Assumption \ref{ass:sigma} and our computational procedure applies. 

The procedure can also be used to implement other likelihood-based methods. For example, it can be combined with data augmentation and Markov chain Monte Carlo methods to implement a Bayesian estimator that can flexibly deal with unobserved heterogeneity.  

Two types of empirical application of the MHT framework can be distinguished. First, it can be used as a descriptive framework, much like \cites{jrssb72:cox} proportional hazards model and \cites{ecta79:lancaster} mixed proportional hazards model. Section \ref{s:strike}'s analysis of \cites{jem85:kennan} strike data shows that estimates of the MHT model have descriptive appeal, with natural interpretations that nicely complement those that could be obtained from a proportional hazards analysis. Indeed, in statistics, there is substantial interest in the descriptive analysis of duration data with first hitting time  models \citep{ss95:singpurwalla,ss97:yashinmanton,ss01:aalengjessing,ss06:leewhitmore}. 

Second, it can be applied to the structural empirical analysis of heterogeneous agents' optimal stopping decisions. \citet{ecta12:abbring} presents a range of examples, based on the type of optimal stopping models that are reviewed and analyzed in \citet{pup94:dixitpindyck,pup09:stokey,springer06:kyprianou,springer07:boyarchenkolevensdorskii}. These include \cites{qje86:mcdonaldsiegel} model for the optimal timing of an irreversible investment;  a model of unemployment durations based on \cites{jpe89:dixit} model of entry and exit, complemented with heterogeneity in transition costs; and a model of job separations with heterogeneous search. The identification results in \citet{ecta12:abbring,are10:abbring} show that data on durations and covariates are informative on the economic primitives of such models. The methods developed in this paper can be applied to measure those primitives.

\section*{Acknowledgements}

We are grateful to Yanqin Fan, the editor (Dennis Kristensen), two referees, and attendees of various conferences and seminars for their comments. We thank Justin Dijk for excellent research assistance. The research of Jaap Abbring is financially supported by the Dutch Research Council (NWO) through Vici grant 453-11-002. Tim Salimans worked on this paper while employed at Erasmus University Rotterdam.

\pdfbookmark[0]{References}{pdfbm:refs}
\bibliographystyle{chicago}
\bibliography{alljaap2,mhte}

\end{document}